\documentclass[final,3p,times,twocolumn]{elsarticle}

\usepackage{amssymb}
\usepackage{amsmath}

\usepackage{xcolor} 
\usepackage{algorithm} 
\usepackage{algpseudocode} 
\usepackage{bm}
\usepackage{upgreek} 

\makeatletter
\def\ps@pprintTitle{%
   \let\@oddhead\@empty
   \let\@evenhead\@empty
   \def\@oddfoot{\footnotesize
   This work has been submitted to a peer-reviewed journal.\hfill}%
   \let\@evenfoot\@oddfoot}
\makeatother

\begin{document}

\begin{frontmatter}

\title{Steerable Invariant Beamformer Using a Differential Line Array of Omnidirectional and Directional Microphones with Null Constraints}
 \author{Yankai Zhang\textsuperscript{a,b,*}}
 \author{Jiafeng Ding\textsuperscript{a,b}}
 \author{Jingjing Ning\textsuperscript{a,b}}
 \author{Qiaoxi Zhu\textsuperscript{c}}
 
 \affiliation[a]{organization={Anhui Digital Intelligent Engineering Research Center for Agricultural Products Quality Safety},
            city={Fuyang},
            postcode={236037},
            state={Anhui},
             country={China}}

 \affiliation[b]{organization={Anhui  Research Center of Generic Technology in Photovoltaic Industry, Fuyang Normal University},
             city={Fuyang},
            postcode={236037},
            state={Anhui},
            country={China}}

 \affiliation[c]{organization={Faculty of Engineering and IT, University of Technology Sydney},
             city={Sydney},
            postcode={NSW 2007},
            country={Australia}}

\begin{abstract}
Line differential microphone arrays have attracted attention for their ability to achieve frequency-invariant beampatterns and high directivity. Recently, the Jacobi-Anger expansion-based approach has enabled the design of fully steerable-invariant differential beamformers for line arrays combining omnidirectional and directional microphones. However, this approach relies on the analytical expression of the ideal beam pattern and the proper selection of truncation order, which is not always practical. This paper introduces a null-constraint-based method for designing frequency- and steerable-invariant differential beamformers using a line array of omnidirectional and directional microphones. The approach employs a multi-constraint optimisation framework, where the reference filter and ideal beam pattern are first determined based on specified nulls and desired direction. Subsequently, the white noise gain constraint is derived from the reference filter, and the beampattern constraint is from the ideal beam pattern. The optimal filter is then obtained by considering constraints related to the beampattern, nulls, and white noise gain. This method achieves a balance between white noise gain and mean square error, allowing robust, frequency- and steerable-invariant differential beamforming performance. It addresses limitations in beam pattern flexibility and truncation errors, offering greater design freedom and improved practical applicability. Simulations and experiments demonstrate that this method outperforms the Jacobi-Anger expansion-based approach in three key aspects: an extended effective range, improved main lobe and null alignment, and greater flexibility in microphone array configuration and beam pattern design, requiring only steering direction and nulls instead of an analytic beam pattern expression.

\end{abstract}




\end{frontmatter}


\section{Introduction}
\label{sec1}

Beamforming has been widely applied in various fields, such as radar \cite{pillai2012array,van2002optimum}, sonar \cite{ma2012theoretical,yan2019broadband}, speech communication \cite{benesty2008microphone,benesty2021array}, and personal sound zone creation\cite{pan2025sparse,zhang2024design1}. Beamforming techniques can be broadly categorised into two types: adaptive beamformers and fixed beamformers, etc. Adaptive beamformers \cite{capon1969high,frost1972algorithm, griffiths1982alternative, cox1987robust} estimate statistics from array data to optimise beamformer coefficients, making them effective in suppressing directional noise and interference. However, they may suffer from signal distortion due to inaccurate estimation of nonstationary signal and noise statistics. In contrast, fixed beamformers avoid these issues by not relying on real-time data. Common fixed beamformers include delay-and-sum \cite{benesty2008microphone}, superdirective \cite{doclo2007superdirective,wang2023general,wang2023robust}, and differential beamformers \cite{elko2000superdirectional,benesty2012study,benesty2015design, zheng2025design, jin2025design}. Differential beamformers, in particular, are notable for their compact size, frequency-invariant beam pattern, and high directivity.

Differential beamformers approximate acoustic pressure differentials using finite differences and can be implemented with various array geometries, such as line array~\cite{benesty2012study,chen2014design,pan2015theoretical,tu2019mainlobe}, circular array~\cite{benesty2015design,huang2017design,huang2018insights,wang2023mode}, and spherical array~\cite{zhao2024differential}. While circular and spherical arrays enable fully steerable beamforming in 2D and 3D spaces, their larger size limits practical applications. Linear differential microphone arrays (LDMAs), on the other hand, are widely used due to their compatibility with devices like smart TVs and personal computers.

LDMAs design methodologies can be classified into four categories: the cascade method \cite{elko2000superdirectional,itzhak2022multistage,zhao2023design}, the null-constrained method \cite{chen2014design,benesty2015design,huang2020continuously,wang2021robust}, the series expansion method \cite{zhao2014design,zhao2016design,huang2018insights,wang2024theoretical}, and the multi-constraint optimisation method \cite{huang2022fundamental,hao2023optimization,xie2024design,bi2025design}. Most research has focused on end-fire designs unsuitable for applications requiring beam steering, such as capturing speech from multiple directions in meetings. Recent work has explored steerable LDMAs. Jin et al. \cite{jin2020steering} derived steering conditions and designed steerable LDMAs with the null-constrained method. Yu \cite{yu2023eigenbeam} developed an eigenbeam-space transformation based steerable differential beamforming for line arrays. Our prior work \cite{zhang2023broadband, zhang2024design} proposed a series expansion based method to design broadside and steerable differential beam patterns using a loudspeaker line array. However, all the above methods are not steerable-invariant, and the directivity factor (DF) still varies with the steering direction.

Recent studies \cite{luo2023design} have shown that line arrays with omnidirectional and bidirectional microphones can achieve full steerability using the Jacobi-Anger series expansion. This approach was extended to linear superarrays, which combine omnidirectional and directional microphones \cite{luo2024design}. Although the improved Jacobi-Anger method can reduce beam pattern errors by increasing the truncation order, this approach inevitably leads to a decrease in White Noise Gain (WNG). thereby compromising the robustness. Furthermore, this method presents several challenges in practical applications: 1) It requires an analytical expression of the desired beam pattern, which limits the flexibility of beamformer design. 2) The expansion error increases with frequency, leading to main lobe deviation and higher sidelobes at high frequencies. 3) Selecting an appropriate truncation order to achieve a balance between WNG and beam pattern errors remains an unresolved issue.

This paper introduces a null-constraint-based method for designing frequency- and steerable-invariant differential beamformers using a line array of omnidirectional and directional microphones. The approach employs a multi-constraint optimisation framework, where the reference filter and ideal beam pattern are first determined based on specified nulls and desired direction. Subsequently, the white noise gain constraint is derived from the reference filter, and the beampattern constraint is from the ideal beam pattern. The optimal filter is then obtained by considering constraints related to the beampattern, nulls, and white noise gain. This method achieves a balance between white noise gain and mean square error, allowing robust, frequency- and steerable-invariant differential beamforming performance. It addresses limitations in beam pattern flexibility and truncation errors, offering greater design freedom and improved practical applicability. Simulations and experiments demonstrate that this method outperforms the Jacobi-Anger expansion-based approach in three key aspects: an extended effective range, improved main lobe and null alignment, and greater flexibility in microphone array configuration and beam pattern design, requiring only steering direction and nulls instead of an analytic beam pattern expression.

\section{Preliminary}
\label{sec2}
\renewcommand{\figurename}{Fig.}
\begin{figure}[htbp]
\centering
\includegraphics[width=0.45\textwidth]{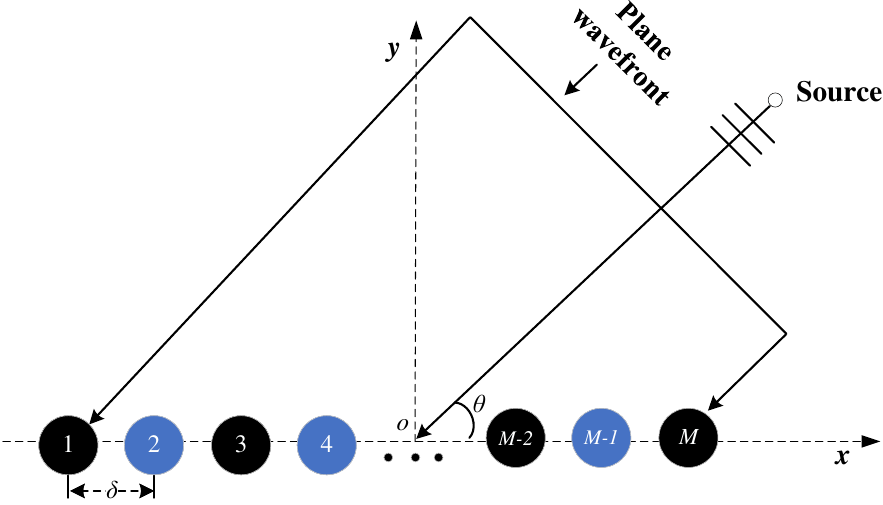}
\caption{A uniform line array with \( M \) microphones: black dots denote omnidirectional microphones, and blue dots denote directional microphones.}
\label{Fig1_schematic}
\end{figure}
\subsection{Signal model for a microphone line array with omni- and directional microphones}
\label{subsec2_1}
A line array of $M$ microphones with spacing $\delta$ is shown in Fig.~\ref{Fig1_schematic}, where omnidirectional and directional microphones are arranged alternately in a uniform pattern. Without loss of generality, we assume that all sensors lie on the $x$-axis, with the line array centred at the origin. The directional pattern of the $m$-th microphone ($m$ = 1, 2,..., $M$) is 
\begin{equation}
  {{g}_{m}}(\theta ,\alpha )={{a}_{m}}+(1-{{a}_{m}})\cos (\theta -\alpha ),
\label{eq:g_m(theta,alpha)}  
\end{equation}
where $\theta$ is the incident angle of sound, and the directional coefficient $a_m$ depends on the type of microphone: omnidirectional ($a_m = 1$), bidirectional ($a_m = 0$), cardioid ($a_m = 0.5$), hypercardioid ($a_m = 1/3$) and supercardioid ($a_m = \sqrt{2} - 1$). In this paper, the microphone's look direction is $\alpha = 90^\circ$, so
\begin{equation}
g_m(\theta) = a_m + (1 - a_m)\sin\theta.
\label{eq:g_m(theta)}
\end{equation}

A plane wave impinges on the line array from direction $\theta$, and the corresponding steering vector (of length $M$) for the line array is 
\begin{equation}
\tilde{\mathbf{d}}(k,\theta) = \mathbf{G}(\theta ) \mathbf{d}(k,\theta ),
\label{eq:tilde_d(k,theta)}
\end{equation}
where $k = 2\pi f / c$ is the wave number with frequency $f$ and the speed of sound, i.e. $c$ = 340 m/s, the directionality of the $M$ microphones forms a diagonal matrix
\begin{equation}
\mathbf{G}(\theta ) = {\rm{diag}}\left[{g_1}(\theta ),{g_2}(\theta ),...,{g_M}(\theta )\right],
\label{eq:G(theta)}
\end{equation}
which is combined with the conventional steering vector
\begin{equation}
\mathbf{d}(k,\theta ) = {\left[ {\begin{array}{*{20}{c}}
{{e^{jk{x_1}\cos \theta }}}& \cdots &{{e^{jk{x_M}\cos \theta }}}
\end{array}} \right]^\text{T}}\ ,
\label{eq:d(k,theta)}
\end{equation}
where $j$ is the imaginary unit, $x_m\, (m=1,\ldots,M)$ represents the position of the $m$-th microphone on the $x$ axis, where ${x_m} =  - (M + 1)\delta /2 + m\delta $, the superscript $^\text{T}$ represents the transpose operator.

For a desired signal from the direction $\theta_s$, the signals received in the frequency domain are 
\begin{equation}
\mathbf{y}(f) = \tilde{\mathbf{d}}(k, \theta_s) x(f) + \mathbf{v}(f),
\label{eq:y(f)}
\end{equation}
where $x(f)$ is the desired signal and $\mathbf{v}(f)$ is the noise vector, where each element corresponds to the noise received by an individual microphone. The dependence on $k$ and $f$ is omitted hereafter for brevity.

\subsection{Steerable beamforming and performance metrics}
\label{subsec2_2}
Beamforming aims to extract the desired signal from noisy observations by applying a complex weighting vector $\mathbf{h}$ to the received signals $\mathbf{y}$
\begin{equation}
\begin{aligned}
z &= \mathbf{h}^\text{H} \mathbf{y} \\
  &= \mathbf{h}^\text{H} \tilde{\mathbf{d}}(\theta_s) x + \mathbf{h}^\text{H} \mathbf{v},
\end{aligned}
\label{eq:z}
\end{equation}
where the superscript $^\text{H}$ is the conjugate transpose. Additionally, a distortionless constraint is applied to promote the preservation of the desired signal
\begin{equation}
\mathbf{h}^\text{H} \tilde{\mathbf{d}}(\theta_s) = 1.
\label{eq:distortionless constraint}
\end{equation}

Three common matrices used to evaluate the performance of the beamformer are the beam pattern, the white noise gain (WNG), and the directivity factor (DF). The beam pattern assesses the spatial response of the beamformer to plane waves from different directions ($\theta$)
\begin{equation}
\mathcal{B}(\mathbf{h}, \theta) = \mathbf{h}^\text{H} \tilde{\mathbf{d}}(\theta).
\label{eq:beampattern}
\end{equation}

The WNG quantifies the beamformer's robustness to noise and microphone mismatches (e.g., in gain, phase, and position), and is defined as
\begin{equation}
\mathcal{W}(\mathbf{h}) = \frac{\left| \mathbf{h}^\text{H} \tilde{\mathbf{d}}(\theta_s) \right|^2}{\mathbf{h}^\text{H} \mathbf{h}}.
\label{eq:WNG}
\end{equation}
When the distortionless constraint (\ref{eq:distortionless constraint}) is satisfied, the WNG and the beamformer's norm are related as
\begin{equation}
||\mathbf{h}||_2^2 = \mathbf{h}^\text{H} \mathbf{h} = \frac{1}{\mathcal{W}(\mathbf{h})}.
\label{eq:h_norm2}
\end{equation}

The DF characterises the beamformer's directional response. The two-dimensional DF, which quantifies spatial gain in a cylindrically isotropic noise field, is defined as
\begin{equation}
\mathcal{D}(\mathbf{h}) = \frac{\left| \mathcal{B}(\mathbf{h}, \theta_s) \right|^2}{\frac{1}{2\pi} \int_0^{2\pi} \left| \mathcal{B}(\mathbf{h}, \theta) \right|^2 d\theta} = \frac{\left| \mathbf{h}^\text{H} \tilde{\mathbf{d}}(\theta_s) \right|^2}{\mathbf{h}^T \tilde{\boldsymbol{\Gamma}} \mathbf{h}} 
\label{eq：DF}
\end{equation}
where
\begin{equation}
\tilde{\boldsymbol{\Gamma}} = \frac{1}{2\pi} \int_0^{2\pi} \tilde{\mathbf{d}}(\theta) \tilde{\mathbf{d}}^\text{H}(\theta) d\theta 
\label{eq:tilde_gamma}
\end{equation}
is an $M \times M$ square matrix. The $(m,n)$-th element (for $m,n=1,2,\ldots,M$) is
\begin{align}
\left[\mathbf{\tilde{\Gamma}}\right]_{m,n} = &\frac{1}{2} \left[ {1 - ({a_m} + {a_n}) + 3{a_m}{a_n}} \right]{J_0}(k{x_m} - k{x_n})  \nonumber\\ 
&+ \frac{1}{2}(1 - {a_m})(1 - {a_n}){J_2}(k{x_m} - k{x_n}).
\label{eq:tilde_gamma_mn}
\end{align}
where $J_n(\cdot)$ ($n=0,2$) 
is the Bessel function of the first kind of order $n$, with the detailed derivation of (\ref{eq:tilde_gamma_mn}) provided in Appendix A.

\subsection{Steerable differential beamforming}
The ideal $N$-th order steerable differential beam pattern \cite{huang2017design}
\begin{equation}
\mathcal{B}_{N,\theta_s}(\theta) = \sum_{n=0}^N \alpha_{N,n} \cos(n\theta - n\theta_s) = \mathbf{a}_N^T \mathbf{c}_{N,\theta_s}(\theta), 
\label{eq:ideal_beampattern}
\end{equation}
where its main lobe is directed at the desired angle $\theta_s$, the real coefficients
\begin{equation}
\mathbf{a}_N = \begin{bmatrix} \alpha_{N,0} & \alpha_{N,1} & \cdots & \alpha_{N,N} \end{bmatrix}^T, 
\label{eq:alpha_N_vec}
\end{equation}
\begin{equation}
\mathbf{c}_{N,\theta_s}(\theta) = \begin{bmatrix} 1 & \cos(\theta - \theta_s) & \cdots & \cos(N\theta - N\theta_s) \end{bmatrix}^T. 
\label{eq:c_N_thetas_vec}
\end{equation}
The ideal beam pattern equals 1 in the desired direction $\theta = \theta_s$, i.e., $\mathcal{B}_{N,\theta_s}( \theta_s) = 1$. Therefore, 
\begin{equation}
\sum_{n=0}^N \alpha_{N,n} = 1.
\label{eq:sigma_alpha_N,n}
\end{equation}

The recently proposed series expansion method \cite{luo2024design} is the first to approximate the ideal beam pattern of a steerable differential beamformer using a microphone line array with omnidirectional and directional microphones. By incorporating directional microphones, the spatial degrees of freedom are extended, enabling the transition from steerable to invariant-steerable differential beamforming. This ensures that the same beam pattern is maintained for different desired directions $\theta_s$, even with a line array. 

However, the method requires an analytical formulation of the ideal beam pattern, which may be limited to certain formulatable forms and not always available for practical applications. To overcome this limitation, we propose a null-constrained design approach that eliminates the need for an exact beam pattern. It relies only on the desired direction and the null directions, simplifying implementation for real-world scenarios. This method becomes even more effective when both the desired and the unwanted/disturbing sound directions are known.

\section{Method}
\subsection{Steerable differential beamforming with the null-constrained (NC) method}
\label{section_4_1}
Assume that the \( N \)th-order differential beamformer, with its main beam directed at \( 0 \), has \( N \) distinct nulls  
\[
0 < \theta'_{N,1} < \cdots < \theta'_{N,N} \leq \pi.
\]
When the main beam shifts to \( \theta_s \), the nulls of the ideal beam pattern after deflection, given by
\[
\theta_{N,n} = \theta_s + \theta'_{N,n}, \quad n=1,\dots,N,
\]
must satisfy  
\[
\theta_s < \theta_s + \theta'_{N,1} < \cdots < \theta_s + \theta'_{N,N} \leq \theta_s + \pi.
\]
Since the ideal beam pattern (\ref{eq:ideal_beampattern}) is an even function and symmetric around the line \( \theta_s \leftrightarrow \pi + \theta_s \), it must also have \( N \) nulls satisfying
\[
\theta_s - \pi \leq \theta_s - \theta'_{N,N} < \cdots < \theta_s - \theta'_{N,1} < \theta_s.
\]
Thus, we obtain
\begin{equation}
\mathcal{B}_{N,\theta_s}(\theta_{N,n}) = \mathcal{B}_{N,\theta_s}(\theta_s \pm \theta'_{N,n}) = 0, \;  n = 1,2,\ldots,N.
\label{eq:nulls_of_ideal_beampattern}
\end{equation}
The null-constrained method approximates the resulting beam pattern to the ideal one by enforcing a gain of 1 in the desired direction and 0 at the nulls. This leads to the following linear system of equations
\begin{equation}
\mathbf{D} \mathbf{h} = \bm{\upgamma}_{2N+1},
\label{eq:nc_method}
\end{equation}
where
\begin{equation}
\mathbf{D} = 
\begin{bmatrix}
\tilde{\mathbf{d}}^\text{H}(\theta_s) \\
\tilde{\mathbf{d}}^\text{H}(\theta_s + \theta'_{N,1}) \\
\tilde{\mathbf{d}}^\text{H}(\theta_s - \theta'_{N,1}) \\
\vdots \\
\tilde{\mathbf{d}}^\text{H}(\theta_s + \theta'_{N,N}) \\
\tilde{\mathbf{d}}^\text{H}(\theta_s - \theta'_{N,N})
\end{bmatrix}.
\label{eq:D}
\end{equation}
is an $(2N+1) \times M$ matrix and 
\begin{equation}
\bm{\upgamma}_{2N+1} = \begin{bmatrix} 1 & 0 & \cdots & 0 \end{bmatrix}^T
\label{eq:gamma_vec_2N+1}
\end{equation}
is a vector of length $2N+1$, whose first element is 1 and all other elements are 0. Two special cases of the method: (1) If the \( N \)th-order differential beamformer has a null opposite to the desired direction (\(\theta_{N,N}' = \pi\)), the two corresponding identical constraints in (\ref{eq:nc_method}) should be merged. (2) If nulls overlap (multiplicity \( > 1 \)), the constraint matrix \( \mathbf{D} \) must be adjusted. See \cite{huang2020continuously} for details.

From (\ref{eq:nc_method}), a steerable-invariant \( N \)th-order differential beamformer requires at least \( 2N+1 \) microphones (\( M \geq 2N+1 \)), corresponding to a line array with at least \( N+1 \) omnidirectional and \( N \) directional microphones, consistent with \cite{luo2024design}. When the minimum requirement \( M = 2N+1 \) is met and \( \mathbf{D} \) is invertible, the solution of (\ref{eq:nc_method}) is
\begin{equation}
\mathbf{h} = \mathbf{D}^{-1} \bm{\upgamma}_{2N+1}.
\label{eq:h_inverse_D}
\end{equation}
When $M > 2N+1$, the beamformer with maximum WNG can be described as
\begin{equation}
\min_{\mathbf{h}}\, \mathbf{h}^\text{H} \mathbf{h} \quad \text{s.t.} \,\,\mathbf{D} \mathbf{h} = \bm{\upgamma}_{2N+1}.
\label{eq:mWNG_problem}
\end{equation}
The solution of (\ref{eq:mWNG_problem}) is the maximum WNG filter
\begin{equation}
\label{eq:h_mWNG}
\mathbf{h}_{\text{mWNG}} = \mathbf{D}^\text{H} (\mathbf{D} \mathbf{D}^\text{H})^{-1} \bm{\upgamma}_{2N+1}.
\end{equation}
Based on (\ref{eq:WNG}), the maximum WNG under the constraints of the desired direction and nulls is
\begin{equation}
\mathcal{W}_{\max} = 10 \log_{10} \frac{1}{\mathbf{h}_{\text{mWNG}}^\text{H} \mathbf{h}_{\text{mWNG}}}.
\label{eq:W_max}
\end{equation}

\subsection{Ideal beam pattern under null constraint}
The solution (\ref{eq:h_mWNG}) constrains only the desired direction and nulls, potentially leading to a suboptimal beam pattern, for example, varying for different desired directions. Considering the desired direction, nulls, and beam pattern symmetry, the coefficients $\mathbf{a}_N$ of the ideal beam pattern (\ref{eq:ideal_beampattern}) are
\begin{equation}
\mathbf{a}_N = \mathbf{C}^{-1} \bm{\upgamma}_{N+1},
\label{eq:calc_alpha_vec}
\end{equation}
where
\begin{equation}
\mathbf{C} = \begin{bmatrix} \mathbf{c}_{N,\theta_s}(\theta_s) & \mathbf{c}_{N,\theta_s}(\theta_{N,1}) & \mathbf{c}_{N,\theta_s}(\theta_{N,2}) & \cdots & \mathbf{c}_{N,\theta_s}(\theta_{N,N}) \end{bmatrix}^T, 
\label{eq:C}
\end{equation}
is an $(N+1) \times (N+1)$ square matrix and $\bm{\upgamma}_{N+1} = \begin{bmatrix} 1 & 0 & \cdots & 0 \end{bmatrix}^T$
is a vector of length $N+1$, whose first element is 1 and all other elements are 0. 

The mean square error (MSE) quantifies the deviation between the resultant beam pattern and the \(N\)-th order ideal differential beam pattern
\begin{equation}
\epsilon_N(\mathbf{h}) = \frac{1}{2\pi} \int_0^{2\pi} \left| \mathcal{B}(\mathbf{h}, \theta) - \mathcal{B}_{N,\theta_s}(\theta) \right|^2 d\theta.
\label{eq:epsilon_N}
\end{equation}
Substituting (\ref{eq:beampattern}) and (\ref{eq:ideal_beampattern}) into (\ref{eq:epsilon_N}), the MSE takes the quadratic form
\begin{equation}
\epsilon_N(\mathbf{h}) = \mathbf{h}^\text{H} \tilde{\mathbf{\Gamma}} \mathbf{h} - \mathbf{h}^\text{H} \mathbf{q} - \mathbf{q}^\text{H} \mathbf{h} + \xi,
\label{eq:epsilon_quadratic}
\end{equation}
where
\begin{equation}
\mathbf{q} = \frac{1}{2\pi} \int_0^{2\pi} \tilde{\mathbf{d}}(\theta) \mathcal{B}_{N,\theta_s}(\theta) d\theta = \mathbf{Q} \mathbf{a}_N, 
\label{eq:q_vec}
\end{equation}
\begin{equation}
\mathbf{Q} = \frac{1}{2\pi} \int_0^{2\pi} \tilde{\mathbf{d}}(\theta) \mathbf{c}_{N,\theta_s}^T(\theta) d\theta, 
\label{eq:Q}
\end{equation}
\begin{equation}
\xi = \frac{1}{2\pi} \int_0^{2\pi} \left| \mathcal{B}_{N,\theta_s}(\theta) \right|^2 d\theta = \mathbf{a}_N^T \bar{\mathbf{C}} \mathbf{a}_N, 
\label{eq:xi}
\end{equation}
\begin{equation}
\bar{\mathbf{C}} = \frac{1}{2\pi} \int_0^{2\pi} \mathbf{c}_{N,\theta_s}(\theta) \mathbf{c}_{N,\theta_s}^T(\theta) d\theta. 
\label{eq:C_bar}
\end{equation}
The matrix \(\tilde{\mathbf{\Gamma}}\), defined in (\ref{eq:tilde_gamma}), is the pseudo-coherence matrix of the cylindrically isotropic noise field. The complex matrix \(\mathbf{Q}\) has elements \((m, n+1)\) for \(m = 1,2,\dots,M\) and \(n = 0,1,\dots,N\)
\begin{align}
\left[\mathbf{Q}\right]_{m,n+1} = &j^n J_n(k x_m) a_m \cos(n \theta_s) - \frac{1-a_m}{2} \sin(n \theta_s)\nonumber\\
&\left[ j^{n+1} J_{n+1}(k x_m) - j^{n-1} J_{n-1}(k x_m) \right].
\label{eq:Q_m_n+1}
\end{align}
$\bar{\mathbf{C}}$ is a diagonal matrix of size $N+1$, whose element is 
\begin{equation}
\left[\mathbf{\bar{C}}\right]_{m,n} = 
\begin{cases} 
1 & m=n=1, \\
0.5 & m=n \neq 1, \quad (m,n =1,\ldots,N+1) . \\
0 & m \neq n,
\end{cases}
\label{eq:C_bar_m_n}
\end{equation}
The derivation of (\ref{eq:Q_m_n+1}) and (\ref{eq:C_bar_m_n}) is in Appendix B.

\subsection{Steerable-invariant differential beamforming with the improved null-constrained (INC) method}
\label{section4_2}
To achieve an invariant steerable differential beam pattern, we minimise the MSE (\ref{eq:epsilon_quadratic}) under the linear constraints (\ref{eq:nc_method}). To enhance robustness, we also impose a WNG constraint. The optimisation problem is formulated as
\begin{align}
\min_{\mathbf{h}} & \quad \mathbf{h}^\text{H} \tilde{\mathbf{\Gamma}} \mathbf{h} - \mathbf{h}^\text{H} \mathbf{q} - \mathbf{q}^\text{H} \mathbf{h} \nonumber \\
\mathrm{s.t.} & \quad \mathbf{D} \mathbf{h} = \bm{\upgamma}_{2N+1}, \nonumber \\
& \quad \mathbf{h}^\text{H} \mathbf{h} \leq 10^{-\zeta_{\text{WNG}}/10}, \label{eq:INC_problem}
\end{align}
where \(\zeta_{\text{WNG}}\) is the preset threshold, defined as  
\begin{equation}  
\zeta_{\text{WNG}} = \mathcal{W}_{\text{max}} - v \leq \mathcal{W}_{\text{max}},  
\label{eq:zeta_WNG}
\end{equation}  
ensuring it does not exceed \(\mathcal{W}_{\text{max}}\). A larger \( v \) may lead to a greater effort in the array to approximate the ideal beam pattern. In this paper, we set $v = 10$, except in Fig.~\ref{fig:diff_zeta}, where we simulate the effect of different values of $v$ with $v = 0, 5, 10,$ and $20$, respectively.

The convex optimisation problem in (\ref{eq:INC_problem}) can be solved using the CVX toolbox \cite{grant2009cvx}. Algorithm 1 summarises the proposed INC method for designing a steerable-invariant differential beamformer with a line array of omni- and directional microphones.

\begin{algorithm}[t]
\caption{Proposed INC Method}
\label{alg:inc}
\begin{algorithmic}[1]
\Require beam pattern order $N$, desired steering angle $\theta_s$, null angles $\{\theta_{N,n}\}_{n=1}^{N}$, frequency $f$,  
microphone directivity coefficients $\{a_m\}_{m=1}^{M}$, and microphone positions $\{x_m\}_{m=1}^{M}$.
\Ensure Optimal beamforming filter $\mathbf{h}$.
\State \textbf{Null Constraint:} Compute the steering matrix $\mathbf{D}$ using (\ref{eq:D}) and obtain the null-constrained reference filter $\mathbf{h}_{\text{mWNG}}$ from (\ref{eq:h_mWNG}).
\State \textbf{White Noise Gain (WNG) Constraint:} Determine the maximum achievable WNG, $\mathcal{W}_{\text{max}}$, using (\ref{eq:W_max}), and set the WNG constraint as  
      $\zeta_{\text{WNG}} = \mathcal{W}_{\text{max}} - v$, where $v \geq 0$ (in dB).
\State \textbf{beam pattern Parameters:} Compute the ideal beam pattern coefficients $\mathbf{a}_N$ using (\ref{eq:calc_alpha_vec}). Construct the pseudo-coherence matrix $\tilde{\mathbf{\Gamma}}$ from (\ref{eq:tilde_gamma}) and the vector $\mathbf{q}$ using (\ref{eq:q_vec}).
\State \textbf{beam pattern Optimization with Null and WNG Constraints:} Solve the convex optimization problem using CVX:  
\[
\begin{aligned}
    &\underset{\mathbf{h}}{\text{min}} && \operatorname{Re}(\mathbf{h}^\text{H} \tilde{\mathbf{\Gamma}} \mathbf{h} - 2 \mathbf{h}^\text{H} \mathbf{q}) \\
    &\text{s.t.} && \mathbf{D} \mathbf{h} = \bm{\upgamma}_{2N+1}, \\
    &           && \operatorname{Re}(\mathbf{h}^\text{H} \mathbf{h}) \leq 10^{-\zeta_{\text{WNG}}/10}
\end{aligned}
\]
\State \Return Optimal filter $\mathbf{h}$.
\end{algorithmic}
\end{algorithm}

\section{Simulation}
This section validates the proposed methods and evaluates beamformer performance through numerical simulations in the 200 Hz–5 kHz range, covering speech frequencies. Parameters are listed below.

Threshold $\zeta_{\text{WNG}}$: We use $v = 10$ in (\ref{eq:zeta_WNG}), except in Sec.~4.2 (Fig.~\ref{fig:diff_zeta}), where we test $v = 0, 5, 10,$ and $20$.  

Desired direction $\theta_s$: Since differential arrays usually target end-fire ($0^\circ$), we use $\theta_s = 90^\circ$ to demonstrate steerability, except in Sec.~4.3 (Figs.~7 and~8), where different $\theta_s$ values are explored.  

Nulls: Fixed nulls are at $\theta_s + 120^\circ$ (first-order) and $\theta_s + 90^\circ$, $\theta_s + 150^\circ$ (second-order). Except in Sec.~4.5 (Figs.~10 and 11), nulls are computed from $\mathbf{a}_N$ \cite{luo2024design} for comparison. Given $\theta_s$ and $\mathbf{a}_N$, the nulls are obtained by solving $\mathbf{a}_N^T \mathbf{c}_{N,\theta_s}(\theta_{N,n}) = 0$ and (\ref{eq:sigma_alpha_N,n}).  

\begin{figure}[t]
    \centering
    \begin{minipage}{0.99\linewidth}
        \includegraphics[width=0.99\linewidth]{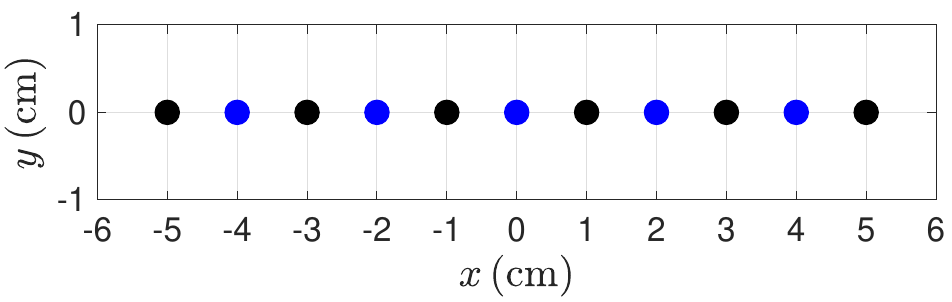}
            \vspace{-0.45cm}
        \caption{Geometry of the uniform line array: black dots denote omnidirectional microphones, blue dots denote directional microphones.} 
        \label{fig:simu_LSA}
    \end{minipage}
\end{figure}

Microphone array: Figure \ref{fig:simu_LSA} shows a uniform line array of 11 microphones—six omnidirectional and five bidirectional. In Sec.~4.4 (Fig.~9), bidirectional microphones are replaced with cardioid, hypercardioid, and supercardioid to examine directivity effects. The spacing is 0.01 m, except in Sec.~4.5 (Figs.~10 and 11), where 0.02 m is used to match \cite{luo2024design}.

\begin{figure}[t]
    \centering
    \begin{minipage}[b]{0.99\linewidth}
        \centering
        \includegraphics[width=0.99\linewidth]{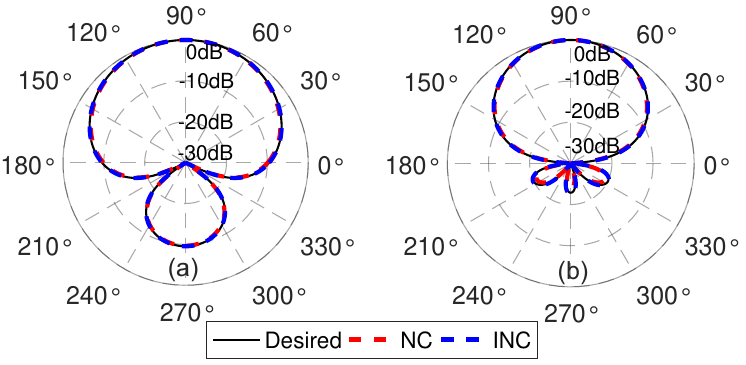}
        \label{fig3a}
     \end{minipage}
     \vspace{-1.2cm}
     \caption{The desired beam pattern and those designed using the proposed NC and INC methods at 500 Hz. (a) First-order differential beam pattern with $\theta_s = 90^\circ$ and a null at $\theta_s + 120^\circ$. (b) Second-order differential beam pattern with $\theta_s = 90^\circ$ and two nulls at $\theta_s + 90^\circ$ and $\theta_s + 150^\circ$.}
     \label{Fig3:NC_INC_compare}
\end{figure}
\begin{figure}[t]
    \centering
    \begin{minipage}[b]{0.99\linewidth}
        \centering
        \includegraphics[width=0.99\linewidth]{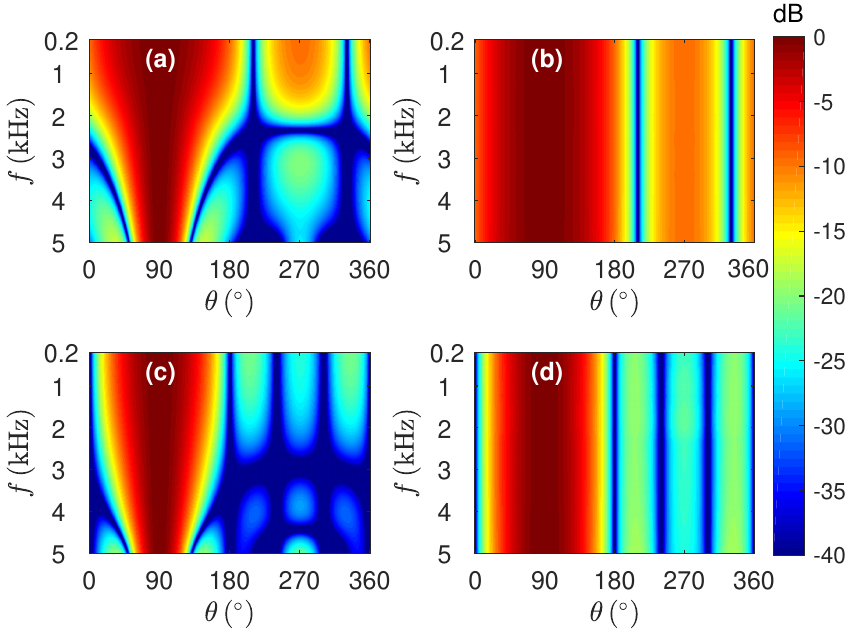}
    \end{minipage}
    \vspace{-0.8cm}
    \caption{Broadband beam patterns for $\theta_s = 90^\circ$ designed using the proposed methods: (a) First-order NC, (b) First-order INC, (c) Second-order NC, and (d) Second-order INC.}
    \label{fig:NC_INC_broadband}
\end{figure}
\begin{figure}[t]
    \centering
    \begin{minipage}[b]{0.99\linewidth}
        \centering
        \includegraphics[width=0.99\linewidth]{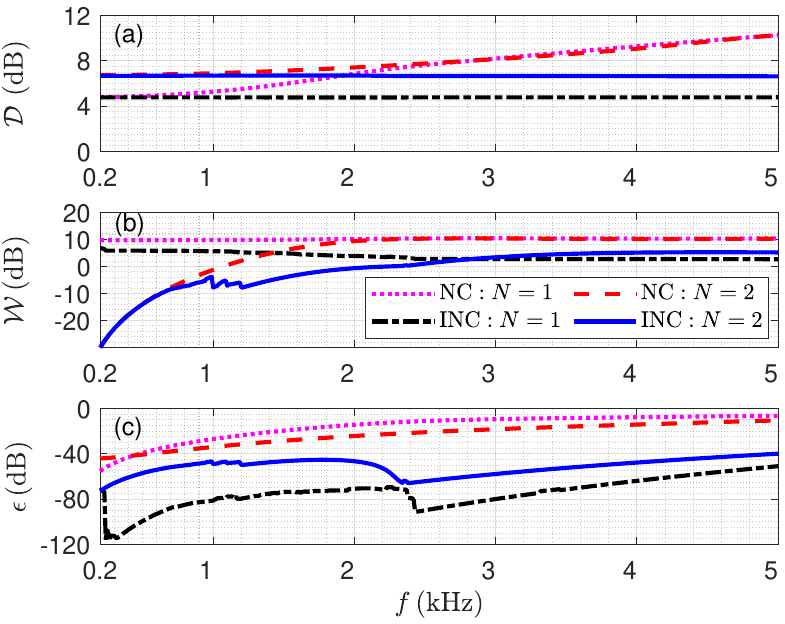}
    \end{minipage}     
    \vspace{-0.8cm}
    \caption{Performance of first-order ($N=1$) and second-order ($N=2$) differential beam patterns designed using the proposed NC and INC methods: (a) directivity factor $\mathcal{D}$, (b) white noise gain $\mathcal{W}$, and (c) mean square error $\epsilon$.}
    \label{fig:NC_INC_performances}
\end{figure}
\begin{figure}[t]
    \centering
    \begin{minipage}[b]{0.99\linewidth}
        \centering
        \includegraphics[width=0.99\linewidth]{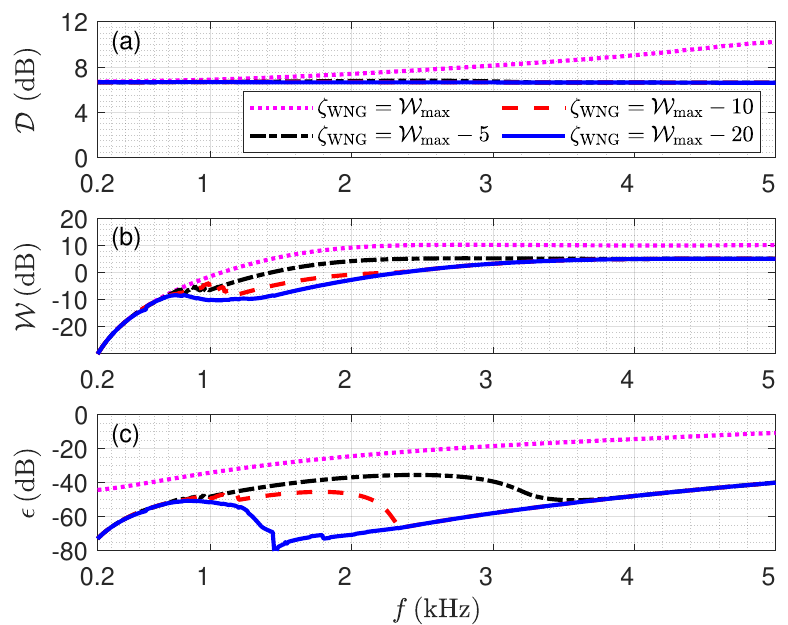}
    \end{minipage}      
    \vspace{-0.8cm}
    \caption{Performance of second-order ($N=2$) differential beam patterns designed using the proposed INC method with different thresholds $\zeta_{\text{WNG}}$: (a) directivity factor $\mathcal{D}$, (b) white noise gain $\mathcal{W}$, and (c) mean square error $\epsilon$.}
    \label{fig:diff_zeta}
\end{figure}

\subsection{Proposed NC and INC methods}
We evaluate the effectiveness of the two null-constrained beamformers proposed in Sections~\ref{section_4_1} and~\ref{section4_2}: the Null-Constrained (NC) method and the Improved Null-Constrained (INC) method.

Figure~\ref{Fig3:NC_INC_compare} shows the desired and resultant beam patterns by the NC and INC methods at 500Hz. In Fig.\ref{Fig3:NC_INC_compare}(a), the first-order differential beam patterns are compared, with the target pattern featuring a main lobe at $\theta_s = 90^\circ$ and a null at $\theta_s + 120^\circ$. Both methods closely replicate the desired pattern. Figure~\ref{Fig3:NC_INC_compare}(b) shows the second-order differential beam patterns, where the main lobe is steered to $\theta_s = 90^\circ$ and nulls are at $\theta_s + 90^\circ$ and $\theta_s + 150^\circ$. The NC and INC methods produce beam patterns nearly identical to the target directivity pattern.

Figure~\ref{fig:NC_INC_broadband} presents the broadband beam patterns from 200 Hz to 5~kHz. The NC method matches the target beam patterns at low frequencies but exhibits a narrowing main lobe as frequency increases. In contrast, the INC method maintains frequency-invariant beam patterns across the entire range.

Figure~\ref{fig:NC_INC_performances}(a) shows that the NC method's directivity factors (DFs) increase with frequency, while the INC method maintains constant DFs, indicating frequency-invariant performance. Figures~\ref{fig:NC_INC_performances}(b) and~\ref{fig:NC_INC_performances}(c) present the broadband white noise gains (WNGs) and mean squared errors (MSEs). The NC method achieves higher WNGs, demonstrating greater robustness, but at the cost of higher MSEs, indicating less accurate approximation of the desired beam patterns. For second-order patterns, WNGs are similar below 700~Hz, but the INC method achieves better MSEs.

This is because the NC method uses all available degrees of freedom to maximize WNG while satisfying linear constraints (\ref{eq:nc_method}). In contrast, the INC method first satisfies these constraints, ensures WNG does not fall below a preset threshold $\zeta_{\text{WNG}}$, and then minimizes MSE. This approach enables a better trade-off between WNG and MSE by adjusting $\zeta_{\text{WNG}}$. 

Given these results, the following sections focus exclusively on the INC method, hereafter referred to as the proposed method.

\subsection{Parameter $\zeta_{\text{WNG}}$}
Figure~\ref{fig:diff_zeta} explores the impact of the parameter $\zeta_{\text{WNG}}$ on the performance of the proposed beamformer, balancing robustness and beam pattern error. The simulation uses the second-order differential beam pattern from Fig.~\ref{Fig3:NC_INC_compare}(b) and considers four $\zeta_{\text{WNG}}$ values: 1) $\mathcal{W}_{\max}$, 2) $\mathcal{W}_{\max} - 5$, 3) $\mathcal{W}_{\max} - 10$, and 4) $\mathcal{W}_{\max} - 20$.

The results show that with $\zeta_{\text{WNG}} = \mathcal{W}_{\max}$, the beamformer mirrors the NC method, with directivity factors (DFs) increasing with frequency. As $\zeta_{\text{WNG}}$ decreases, DFs become more consistent across frequencies. Lower $\zeta_{\text{WNG}}$ values reduce WNGs, particularly in the mid-frequency range, but also decrease beam pattern errors, as more degrees of freedom are allocated to error minimization.

Figures~\ref{fig:diff_zeta}(b) and (c) reveal that $\zeta_{\text{WNG}} = \mathcal{W}_{\max} - 10$ offers the best trade-off, maintaining MSEs below $-40$~dB—an error level sufficiently low for practical applications \cite{wang2023mode}.

\subsection{Steerability}
\begin{figure}[t]
       \centering
    \begin{minipage}[b]{0.99\linewidth}
        \centering
        \includegraphics[width=0.99\linewidth]{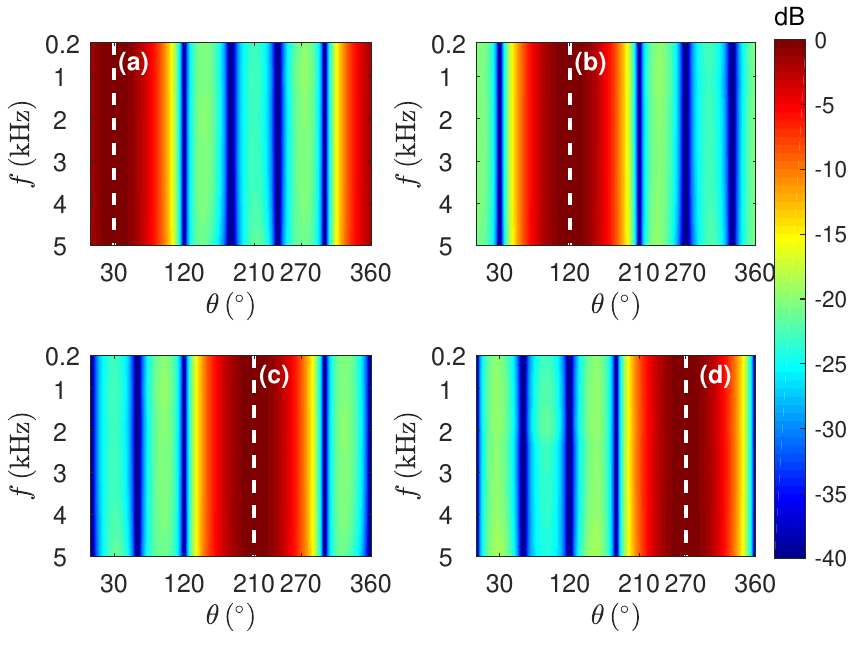}  
    \end{minipage}
    \vspace{-0.8cm}
    \caption{Broadband second-order ($N = 2$) differential beam patterns designed using the proposed INC method for (a) $\theta_s = 30^\circ$, (b) $\theta_s = 120^\circ$, (c) $\theta_s = 210^\circ$, and (d) $\theta_s = 270^\circ$.}
    \label{fig:steerability}
\end{figure}
\begin{figure}[t]
    \centering
    \begin{minipage}[b]{0.99\linewidth}
        \centering
        \includegraphics[width=0.99\linewidth]{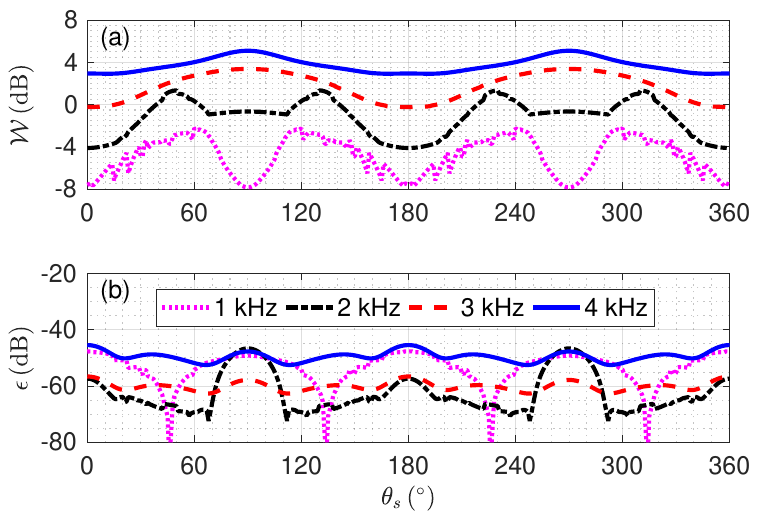}
    \end{minipage}
    \vspace{-0.8cm}
    \caption{Performance of second-order ($N=2$) differential beam patterns designed using the proposed INC method at different frequencies versus desired directions $\theta_s $. (a) white noise gain $\mathcal{W}$, and (b) mean square error $\epsilon$.}
    \label{fig:diff_thetas}
\end{figure}

Figures~\ref{fig:steerability} and~\ref{fig:diff_thetas} examine the steering flexibility of the proposed method using a line array to design a second-order differential beam pattern with the main lobe steered to $\theta_s = 30^\circ$, $120^\circ$, $210^\circ$, and $270^\circ$, and nulls at $\theta_s + 90^\circ$ and $\theta_s + 150^\circ$. Figure~\ref{fig:steerability} shows that the method maintains frequency-invariant beam patterns across frequencies, with the pattern shifting as $\theta_s$ increases, demonstrating full steerability. Additionally, Fig.~\ref{fig:diff_thetas} plots the WNGs and MSEs at 1, 2, 3, and 4~kHz, where WNGs increase with frequency and vary with steering direction, while MSEs stay below $-40$~dB, confirming the method’s ability to design steerable, invariant differential beamformers.

\subsection{Types of Directional Microphones}
\begin{figure}[t]
    \centering
    \begin{minipage}[b]{0.99\linewidth}
        \centering
        \includegraphics[width=0.99\linewidth]{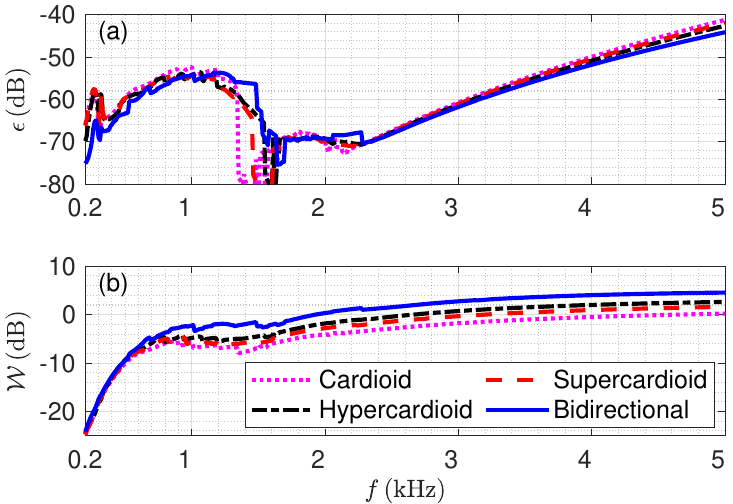}
    \end{minipage}
    \vspace{-0.8cm}
    \caption{Performance of second-order ($N=2$) differential beam patterns designed using the proposed INC method with different types of directional microphones. (a) mean square error $\epsilon$, and (b) white noise gain $\mathcal{W}$.}
    \label{fig:diff_mic_types}
\end{figure}

Figure~\ref{fig:diff_mic_types} explores the impact of different directional microphones on the proposed beamformer performance to achieve a second-order differential beam pattern with a steering angle of \(\theta_s = 60^\circ\) and nulls at \(\theta_s + 90^\circ\) and \(\theta_s + 150^\circ\). The line array (as shown in Fig.~\ref{fig:simu_LSA}) consists of omnidirectional microphones and directional microphones of four types, respectively: 1) Cardioid, 2) Hypercardioid, 3) Supercardioid, and 4) Bidirectional. Figure~\ref{fig:diff_mic_types}(a) shows MSEs below $-40$ dB for all arrays, indicating negligible beam pattern errors. However, as shown in Figure~\ref{fig:diff_mic_types}(b), microphone type significantly affects robustness, particularly at frequencies above 600~Hz, with the bidirectional microphone array demonstrating the highest robustness, consistent with \cite{luo2024design}.

\subsection{Comparison with the Jacobi-Anger Expansion-based Method~\cite{luo2024design}}
\label{sec:jacobi_compare}

\begin{figure}[t]
    \centering
    \begin{minipage}[b]{0.99\linewidth}
        \centering
        \includegraphics[width=0.99\linewidth]{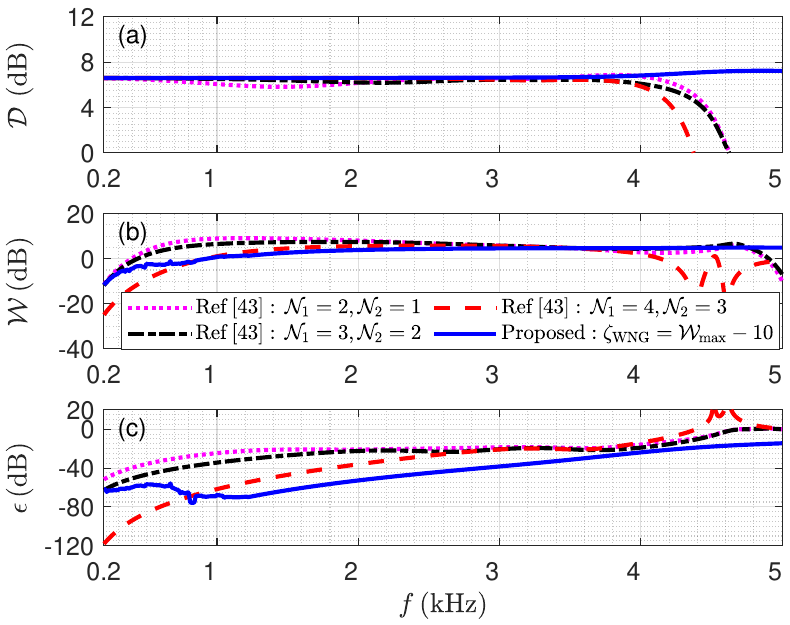}
    \end{minipage}   
    \vspace{-0.8cm}
    \caption{Comparison of the Jacobi-Anger expansion method \cite{luo2024design} with truncation orders (\(\mathcal{N}_1, \mathcal{N}_2\)) = (2,1), (3,2), and (4,3), and the proposed INC method in approximating the second-order cardioid beam pattern at \( \theta_s = 60^\circ \): (a) directivity factor \( \mathcal{D} \), (b) white noise gain \( \mathcal{W} \), and (c) mean square error \( \epsilon \).}
    \label{fig:compare_jacobi}
\end{figure}
\begin{figure}[htbp]
       \centering
    \begin{minipage}[b]{0.99\linewidth}
        \centering
        \includegraphics[width=0.99\linewidth]{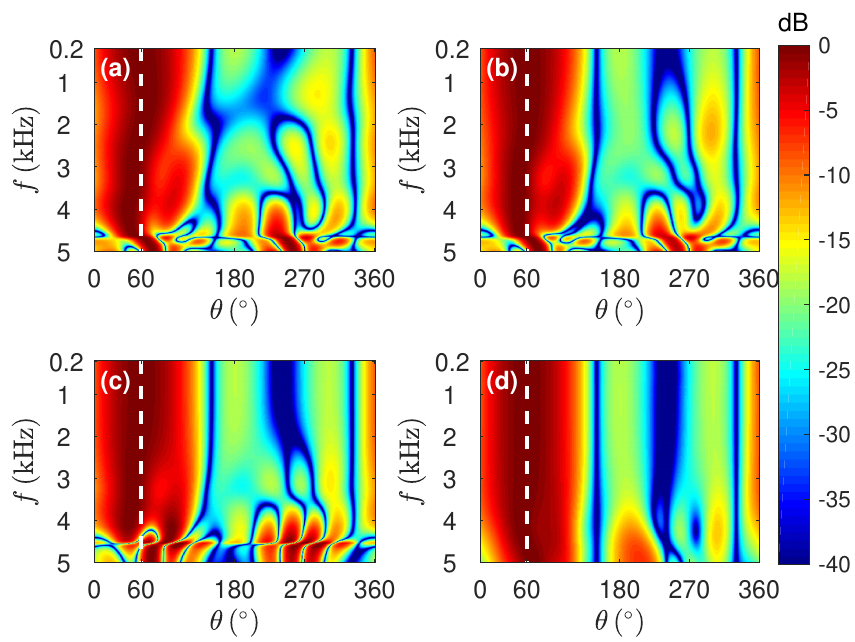}  
    \end{minipage}
    \vspace{-0.8cm}
    \caption{Broadband beam patterns for second-order cardioid approximation using: (a–c) the Jacobi-Anger expansion method \cite{luo2024design} with truncation orders (\(\mathcal{N}_1, \mathcal{N}_2\)) = (2,1), (3,2), and (4,3), respectively; and (d) the proposed INC method. The white dashed line indicates the desired direction \(\theta_s = 60^\circ\).} 
    \label{fig:compare_jacobi_broadband}
\end{figure}

The Jacobi-Anger expansion method \cite{luo2024design} designs fully steerable differential beamformers using line array of omni- and directional microphones. It decomposes the desired beam pattern into components achievable by omnidirectional and directional arrays, approximating them via Jacobi-Anger expansion. Truncation orders $\mathcal{N}_1$ and $\mathcal{N}_2$ control the trade-off between WNG and MSE performance. To highlight the advantages of the proposed method, we compare it with the Jacobi-Anger method. The simulation uses a line array (similar to Fig.~\ref{fig:simu_LSA}) but 0.02 m spacing and bidirectional microphones. The target is a second-order cardioid beam pattern $(\alpha_{2,0}=1/4,\alpha_{2,1}=1/2,\alpha_{2,2}=1/4)$ steered to \(\theta_s = 60^\circ\), with nulls at \(\theta_s + 90^\circ\) and \(\theta_s + 180^\circ\).

Figure~\ref{fig:compare_jacobi} presents the broadband DFs, WNGs, and MSEs. In Fig.~\ref{fig:compare_jacobi}(a), the Jacobi-Anger method shows fluctuating DFs below 4 kHz and rapid degradation above, indicating high-frequency distortion. The proposed method maintains a stable DF below 3 kHz, with a slight increase thereafter. Fig.~\ref{fig:compare_jacobi}(b) shows that the Jacobi-Anger method (with $\mathcal{N}_1 = 4$, $\mathcal{N}_2 = 3$) has the lowest WNGs at low (below 800 Hz) and high (above 4 kHz) frequencies, indicating poor robustness. Although the proposed method's WNGs are lower above 1 kHz, they remain above 0 dB, ensuring practical robustness. Fig.~\ref{fig:compare_jacobi}(c) shows MSEs below $-40$~dB for the proposed method up to 3 kHz, with minimal degradation above, outperforming the Jacobi-Anger method~\cite{luo2024design}. Figure~\ref{fig:compare_jacobi_broadband} further compares the broadband beam patterns. While the Jacobi-Anger method improves frequency invariance with higher truncation orders, it suffers from main lobe shifts and increased sidelobes above 4 kHz. The proposed method maintains an invariant beam pattern below 3 kHz, and despite sidelobes above 4 kHz, the main lobe remains aligned due to the distortionless constraint.

\section{Experiment}
\begin{figure}[t]
    \centering
    \begin{minipage}[b]{0.99\linewidth}
        \centering
        \includegraphics[width=0.99\linewidth]{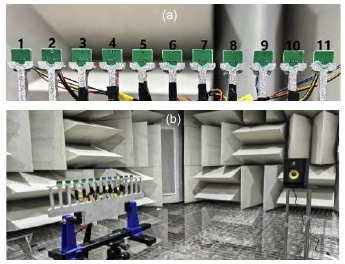}
        \vspace{-0.8cm}
    \caption{Experimental setup: (a) an 11-element uniform line array with odd-numbered omnidirectional microphones and even-numbered bidirectional microphones; (b) the array placed on a turntable with a loudspeaker positioned 3 meters away in an anechoic chamber.}
        \label{fig：exp}
    \end{minipage}
\end{figure}
\begin{figure}[t]
    \centering
    \begin{minipage}{0.99\linewidth}
        \includegraphics[width=0.99\linewidth]{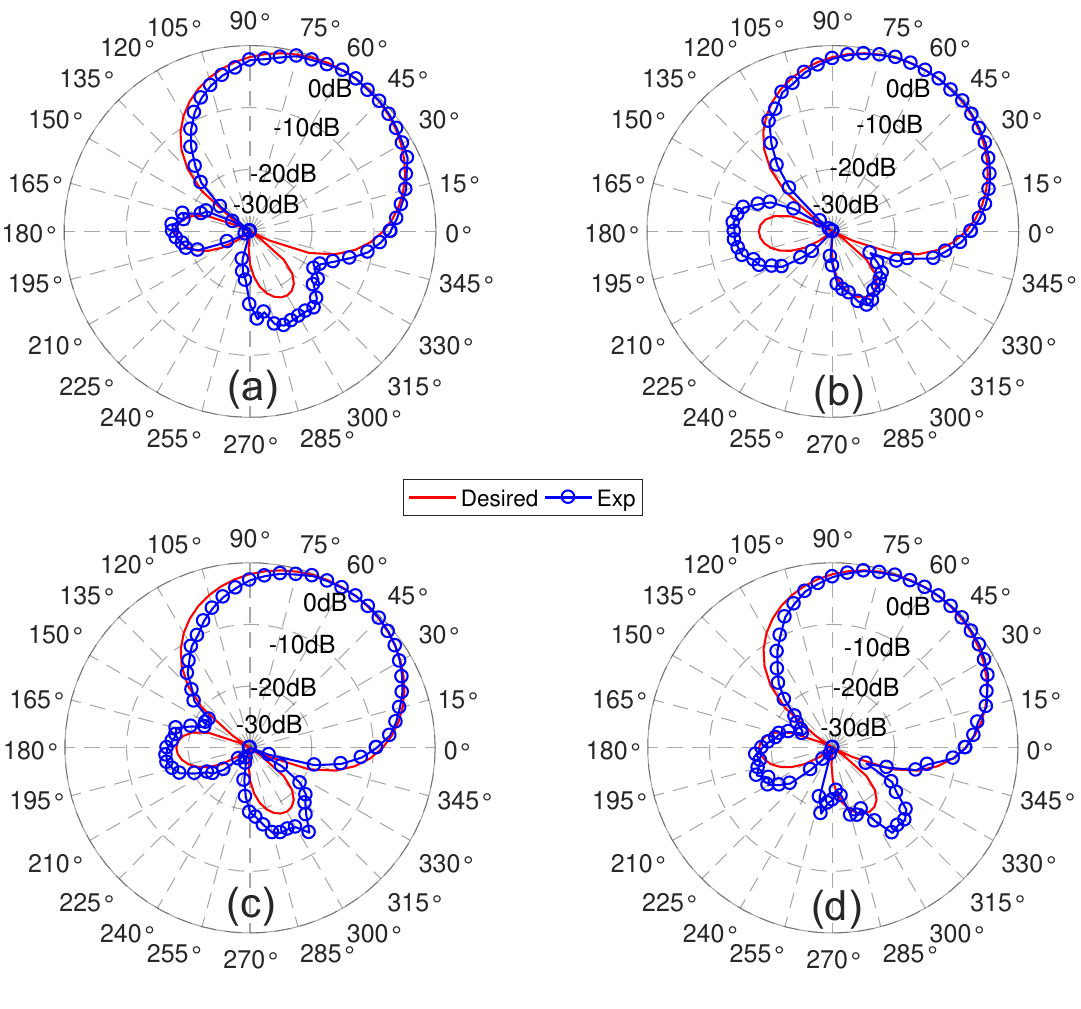}
    \end{minipage}
    \vspace{-0.5cm}
    \caption{Offline and desired beam patterns for second-order cardioid approximation with $\theta_s = 60^\circ$ at different frequencies: (a) 500 Hz, (b) 1 kHz, (c) 2 kHz, and (d) 3 kHz.}
    \label{fig:exp1_single}
\end{figure}
\begin{figure}[t]
    \centering
    \begin{minipage}[b]{0.99\linewidth}
        \centering
        \includegraphics[width=0.99\linewidth]{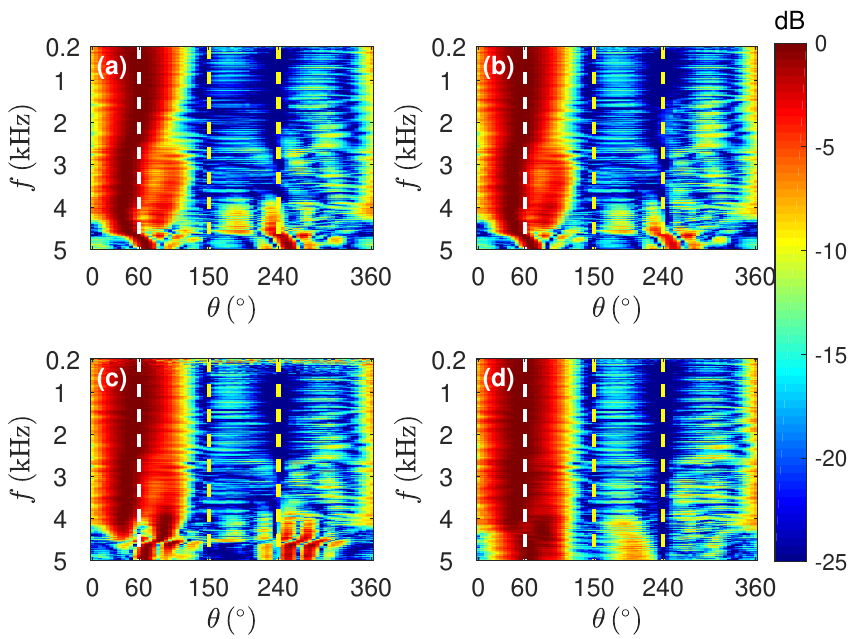}  
    \end{minipage}
    \caption{Offline broadband beam patterns for second-order cardioid approximation using: (a–c) the Jacobi-Anger expansion method \cite{luo2024design} with truncation orders (\(\mathcal{N}_1, \mathcal{N}_2\)) = (2,1), (3,2), and (4,3), respectively; and (d) the proposed INC method. The white dashed line indicates the desired direction \(\theta_s = 60^\circ\), and the yellow dashed lines denote the nulls.} 
    \label{fig:exp2_broadband}
\end{figure}
\begin{figure}[t]
    \centering
    \begin{minipage}[b]{0.99\linewidth}
        \centering
        \includegraphics[width=0.99\linewidth]{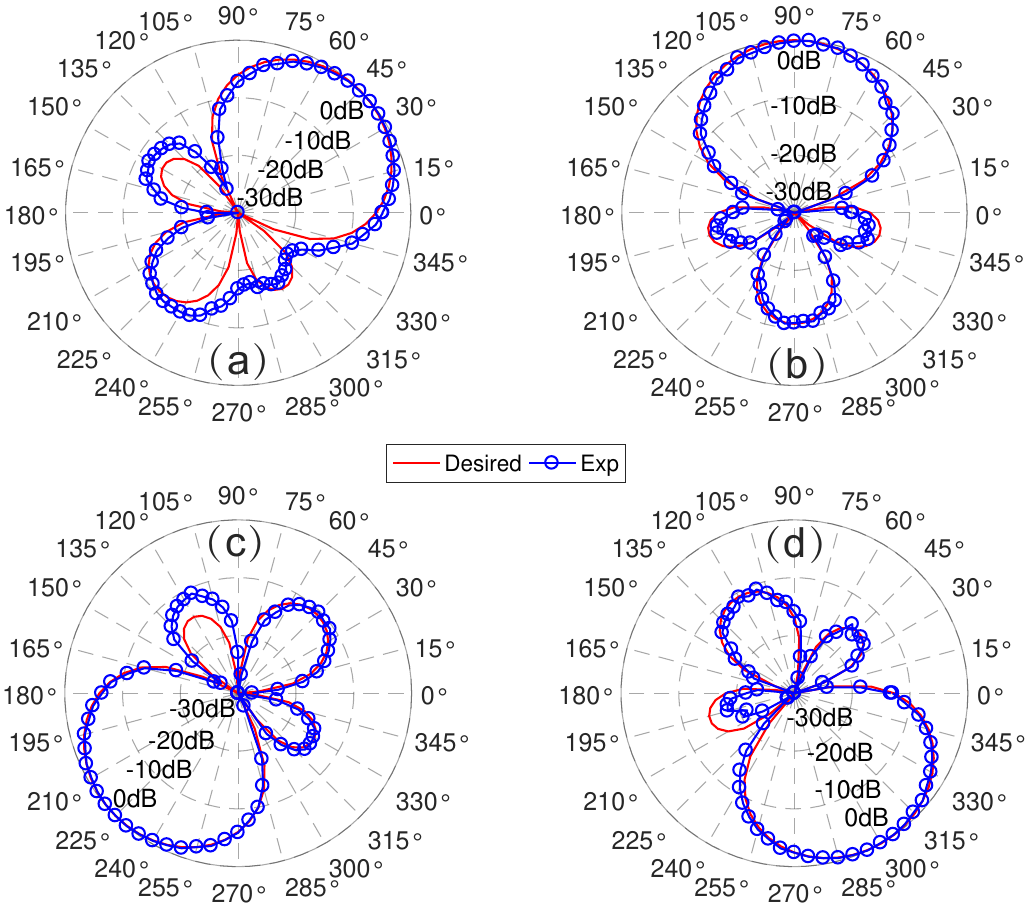}
    \end{minipage}  
    \caption{Offline and desired beam patterns of the second-order differential beamformer, at 1 kHz, for different desired directions: (a) $\theta_s = 45^\circ$, (b) $\theta_s = 90^\circ$, (c) $\theta_s = 225^\circ$, and (d) $\theta_s = 300^\circ$. Conditions: two nulls at $\theta_s + 90^\circ$ and $\theta_s + 135^\circ$.}
    \label{fig:exp3_steerable}
\end{figure}

Experiments were conducted in an anechoic chamber (9.0~m $\times$ 9.6~m $\times$ 5.1~m) to validate the proposed method using an 11-element line array, as shown in Fig.~\ref{fig：exp}(a). The array, consisting of 6 omnidirectional microphones (SPW0878LR5H-1, Knowles, USA) and 5 bidirectional microphones (SKR0600, Soundskrit, Canada) spaced 0.02~m apart, was mounted on a turntable. A loudspeaker (Rokit7, KRK, USA) was placed 3 m from the array's centre at a height of 1.6 m, as shown in Fig.~\ref{fig：exp}(b).

A frequency sweep (20 Hz to 20 kHz) generated by a computer was sent to an audio interface (Fireface UFX III, RME, Germany) and reproduced by the loudspeaker. The acoustic signals received by the line array were transferred to the computer via the same interface to obtain the transfer function between the loudspeaker and the array. This process was repeated for array rotations from 0° to 360° in 5° increments, resulting in 73 measurement points. The transfer functions were used to calculate the offline beam pattern to evaluate the proposed beamformer.

Figure~\ref{fig:exp1_single} shows the performance of approximating a second-order cardioid with $\theta_s = 60^\circ$ (as described in Sec.~\ref{sec:jacobi_compare}) and $\zeta_{\text{WNG}} = \mathcal{W}_{\max} - 10$. The offline and desired beam patterns at 500 Hz, 1 kHz, 2 kHz, and 3 kHz are compared, demonstrating close alignment with minor discrepancies due to measurement errors and array imperfections.

Figure~\ref{fig:exp2_broadband} compares broadband beam patterns designed using the Jacobi-Anger method~\cite{luo2024design} (at various truncation orders) and the proposed INC method. Figures~\ref{fig:exp2_broadband}(a) and (b) show that the Jacobi-Anger method with low truncation orders produces nearly frequency-invariant beam patterns below 2 kHz, but deviates above 2 kHz with large sidelobes above 4 kHz, as seen in Sec.~\ref{sec:jacobi_compare}. Fig.~\ref{fig:exp2_broadband}(c) shows that high truncation orders ($\mathcal{N}_1 = 4$, $\mathcal{N}_2 = 3$) suffer from poor anti-perturbation below 400 Hz and increased sidelobes above 4 kHz. In contrast, the proposed method maintains a frequency-invariant beam pattern up to 3 kHz, with the main lobe and nulls aligning with the ideal pattern even beyond 4 kHz.

To demonstrate the steering flexibility and design convenience of the proposed method, a second-order differential beamformer was designed with the main lobe steered to $\theta_s = 45^\circ$, $90^\circ$, $225^\circ$, and $300^\circ$, with nulls at $\theta_s + 75^\circ$ and $\theta_s + 135^\circ$. The offline beam patterns at 1 kHz, shown in Fig.~\ref{fig:exp3_steerable}, closely match the desired patterns, confirming that the proposed method can design steerable-invariant differential beamformers using omni- and directional microphones, requiring just the desired direction and nulls.

\section{Conclusion}
The proposed null-constraint-based method offers a robust and flexible solution for designing frequency- and steerable-invariant differential beamformers. By employing a multi-constraint optimisation framework, the method effectively balances white noise gain and mean square error, enhancing beamforming performance across various configurations. It overcomes limitations in beam pattern flexibility and truncation errors, allowing greater design freedom and practical applicability. Simulation and experimental results confirm that this approach outperforms the Jacobi-Anger expansion-based method by extending the effective range, improving main lobe and null alignment, and increasing flexibility in microphone array configuration and beam pattern design. This method simplifies the design process by requiring only the steering direction and nulls, enabling more adaptable and efficient implementations of steerable-invariant differential beamformers using a line array.

\appendix
\section{Derivation of Equation (\ref{eq:tilde_gamma_mn})}
\label{app1}
By substituting (\ref{eq:tilde_d(k,theta)}) into (\ref{eq:tilde_gamma}), 
\begin{equation}
\tilde{\mathbf{\Gamma}} = \frac{1}{2\pi} \int_0^{2\pi} \mathbf{G}(\theta) \mathbf{d}(\theta) \mathbf{d}^\text{H}(\theta) \mathbf{G}^\text{H}(\theta) d\theta.
\label{eq:A.1}
\end{equation}
The $(m,n)$-th element (for $m, n = 1, 2, \ldots, M$) is
\begin{equation}
\begin{aligned}
\left[\mathbf{\tilde{\Gamma}}\right]_{m, n} &= \left[\mathbf{\tilde{\Gamma}}\right]_{m,n}^\mathrm{I} + \left[\mathbf{\tilde{\Gamma}}\right]_{m,n}^\mathrm{II} + \left[\mathbf{\tilde{\Gamma}}\right]_{m,n}^\mathrm{III}, \\
\left[\mathbf{\tilde{\Gamma}}\right]_{m,n}^\mathrm{I} &= \frac{{a_m}{a_n}}{{2\pi }}\int_0^{2\pi } {{e^{jk({x_m} - {x_n})\cos \theta }}} d\theta, \\
\left[\mathbf{\tilde{\Gamma}}\right]_{m,n}^\mathrm{II} &= \frac{\left( {{a_m} + {a_n} - 2{a_m}{a_n}} \right)}{{2\pi }}\int_0^{2\pi } {{e^{jk({x_m} - {x_n})\cos \theta }}} \sin \theta d\theta, \\
\left[\mathbf{\tilde{\Gamma}}\right]_{m,n}^\mathrm{III} &= \frac{\left( {1 - {a_m} - {a_n} + {a_m}{a_n}} \right)}{{2\pi }}\int_0^{2\pi } {{e^{jk({x_m} - {x_n})\cos \theta }}} {\sin ^2}\theta d\theta.
\end{aligned}
\label{eq:A.2}
\end{equation}
From the Jacobi-Anger series expansion
\begin{equation}
e^{jk(x_m - x_n) \cos \theta} = \sum_{{n'} = -\infty}^{+\infty} j^{{n'}} J_n(k x_m - k x_n) e^{j{n'} \theta},
\label{eq:A.3}
\end{equation}
the orthogonality of the circular harmonics
\begin{equation}
\frac{1}{2\pi} \int_0^{2\pi} e^{j \tilde{n} \theta} d\theta = 
\begin{cases} 
1, & \tilde{n} = 0, \\
0, & \tilde{n} \neq 0,
\end{cases}
\label{eq:A.4}
\end{equation}
and the Euler's formula,
\begin{equation}
e^{j \theta} = \cos \theta + j \sin \theta,
\label{eq:A.5}
\end{equation}
we have
\begin{equation}
\begin{aligned}
    &\frac{1}{2\pi} \int_0^{2\pi} e^{jk(x_m - x_n) \cos \theta} d\theta = J_0(k x_m - k x_n), \\
    &\frac{1}{2\pi} \int_0^{2\pi} e^{jk(x_m - x_n) \cos \theta} \sin \theta d\theta = 0, \\
    &\frac{1}{2\pi} \int_0^{2\pi} e^{jk(x_m - x_n) \cos \theta} \sin^2 \theta d\theta =\\ & \frac{1}{2} \left[ J_0(k x_m - k x_n) + J_2(k x_m - k x_n) \right],
\end{aligned}
\label{eq:A.6}
\end{equation}
where $J_n(\cdot)$ is the Bessel function of the first kind of order $n$. 
Equation (\ref{eq:tilde_gamma_mn}) is obtained by substituting (\ref{eq:A.6}) into (\ref{eq:A.2}).

\section{Derivation of Equations (\ref{eq:Q_m_n+1}) and (\ref{eq:C_bar_m_n})}
\label{app2}
By substituting (\ref{eq:tilde_d(k,theta)}) and (\ref{eq:c_N_thetas_vec}) into (\ref{eq:Q}), the $(m, n+1)$-th element of $\mathbf{Q}$ (for $m = 1, 2, \ldots, M$ and $n = 0, 1, \ldots, N$)
\begin{equation}
\begin{aligned}
    & \left[\mathbf{Q}\right]_{m,n+1} = \left[\mathbf{Q}\right]^\text{I}_{m,n+1} +\left[\mathbf{Q}\right]^\text{II}_{m,n+1} + \left[\mathbf{Q}\right]^\text{III}_{m,n+1} + \left[\mathbf{Q}\right]^\text{IV}_{m,n+1}, \\
    & \left[\mathbf{Q}\right]^{\text{I}}_{m,n+1} = \frac{a_m \cos(n \theta_s)}{2\pi} \int_{0}^{2\pi} e^{j k x_m \cos \theta} \cos(n \theta) d\theta, \\
    & \left[\mathbf{Q}\right]^{\text{II}}_{m,n+1} = \frac{a_m \sin(n \theta_s)}{2\pi} \int_{0}^{2\pi} e^{j k x_m \cos \theta} \sin(n \theta) d\theta, \\
    & \left[\mathbf{Q}\right]^{\text{III}}_{m,n+1} = \frac{(1 - a_m) \cos(n \theta_s)}{2\pi} \int_{0}^{2\pi} e^{j k x_m \cos \theta} \sin \theta \cos(n \theta) d\theta, \\
    & \left[\mathbf{Q}\right]^{\text{IV}}_{m,n+1} = \frac{(1 - a_m) \sin(n \theta_s)}{2\pi} \int_{0}^{2\pi} e^{j k x_m \cos \theta} \sin \theta \sin(n \theta) d\theta.
\end{aligned}
\label{eq:B.1}
\end{equation}
From (\ref{eq:A.3}) and (\ref{eq:A.5}), we have
\begin{equation}
\begin{aligned}
    &\frac{1}{2\pi} \int_{0}^{2\pi} e^{j k x_m \cos \theta} \cos(n \theta) d\theta = j^n J_n(k x_m), \\
    &\frac{1}{2\pi} \int_{0}^{2\pi} e^{j k x_m \cos \theta} \sin(n \theta) d\theta = 0, \\
    &\frac{1}{2\pi} \int_{0}^{2\pi} e^{j k x_m \cos \theta} \sin \theta \cos(n \theta) d\theta = 0, \\
    &\frac{1}{2\pi} \int_{0}^{2\pi} e^{j k x_m \cos \theta} \sin \theta \sin(n \theta) d\theta =\\& -\frac{1}{2} \left[ j^{n+1} J_{n+1}(k x_m) - j^{n-1} J_{n-1}(k x_m) \right].
\end{aligned}
\label{eq:B.2}
\end{equation}
By combining (\ref{eq:B.1}) and (\ref{eq:B.2}), we obtain Equation (\ref{eq:Q_m_n+1}).
%
%
%
By substituting (\ref{eq:c_N_thetas_vec}) into (\ref{eq:C_bar}), the $(m, n)$-th element of $\mathbf{\bar{C}}$ (for $m = 1,\ldots, N+1$ and $n = 1,\ldots, N+1$) is 
\begin{equation}
\left[\mathbf{\bar{C}}\right]_{m,n} = \frac{1}{2\pi} \int_{0}^{2\pi} \cos(m \theta - m \theta_s) \cos(n \theta - n \theta_s) d\theta.
\label{B.4}
\end{equation}
We categorize the cases according to $m$ and $n$, resulting in (\ref{eq:C_bar_m_n}).
%

  \bibliographystyle{elsarticle-num-names} 
  \bibliography{bibliography_aa2025}

\begin{thebibliography}{45}
\expandafter\ifx\csname natexlab\endcsname\relax\def\natexlab#1{#1}\fi
\providecommand{\url}[1]{\texttt{#1}}
\providecommand{\href}[2]{#2}
\providecommand{\path}[1]{#1}
\providecommand{\DOIprefix}{doi:}
\providecommand{\ArXivprefix}{arXiv:}
\providecommand{\URLprefix}{URL: }
\providecommand{\Pubmedprefix}{pmid:}
\providecommand{\doi}[1]{\href{http://dx.doi.org/#1}{\path{#1}}}
\providecommand{\Pubmed}[1]{\href{pmid:#1}{\path{#1}}}
\providecommand{\bibinfo}[2]{#2}
\ifx\xfnm\relax \def\xfnm[#1]{\unskip,\space#1}\fi
\bibitem[{Pillai(2012)}]{pillai2012array}
\bibinfo{author}{S.~U. Pillai}, \bibinfo{title}{Array signal processing}, \bibinfo{publisher}{Springer Science \& Business Media}, \bibinfo{year}{2012}.
\bibitem[{Van~Trees(2002)}]{van2002optimum}
\bibinfo{author}{H.~L. Van~Trees}, \bibinfo{title}{Optimum array processing: Part IV of detection, estimation, and modulation theory}, \bibinfo{publisher}{John Wiley \& Sons}, \bibinfo{year}{2002}.
\bibitem[{Ma et~al.(2012)Ma, Yang, He, Yang, Sun, and Wang}]{ma2012theoretical}
\bibinfo{author}{Y.~Ma}, \bibinfo{author}{Y.~Yang}, \bibinfo{author}{Z.~He}, \bibinfo{author}{K.~Yang}, \bibinfo{author}{C.~Sun}, \bibinfo{author}{Y.~Wang},
\newblock \bibinfo{title}{Theoretical and practical solutions for high-order superdirectivity of circular sensor arrays},
\newblock \bibinfo{journal}{IEEE Transactions on Industrial Electronics} \bibinfo{volume}{60} (\bibinfo{year}{2012}) \bibinfo{pages}{203--209}.
\bibitem[{Yan(2019)}]{yan2019broadband}
\bibinfo{author}{S.~Yan}, \bibinfo{title}{Broadband array processing}, volume~\bibinfo{volume}{17}, \bibinfo{publisher}{Springer}, \bibinfo{year}{2019}.
\bibitem[{Benesty(2008)}]{benesty2008microphone}
\bibinfo{author}{J.~Benesty}, \bibinfo{title}{Microphone array signal processing}, \bibinfo{publisher}{Springer Verlag}, \bibinfo{year}{2008}.
\bibitem[{Benesty et~al.(2021)Benesty, Cohen, and Chen}]{benesty2021array}
\bibinfo{author}{J.~Benesty}, \bibinfo{author}{I.~Cohen}, \bibinfo{author}{J.~Chen}, \bibinfo{title}{Array beamforming with linear difference equations}, volume~\bibinfo{volume}{20}, \bibinfo{publisher}{Springer}, \bibinfo{year}{2021}.
\bibitem[{Pan et~al.(2025)Pan, Chen, Han, Feng, and Shen}]{pan2025sparse}
\bibinfo{author}{K.~Pan}, \bibinfo{author}{S.~Chen}, \bibinfo{author}{Y.~Han}, \bibinfo{author}{X.~Feng}, \bibinfo{author}{Y.~Shen},
\newblock \bibinfo{title}{Sparse loudspeaker array design for wideband frequency-invariant beamforming with multiple targets},
\newblock \bibinfo{journal}{The Journal of the Acoustical Society of America} \bibinfo{volume}{157} (\bibinfo{year}{2025}) \bibinfo{pages}{369--381}.
\bibitem[{Zhang et~al.(2024)Zhang, Wei, and Zhu}]{zhang2024design1}
\bibinfo{author}{Y.~Zhang}, \bibinfo{author}{H.~Wei}, \bibinfo{author}{Q.~Zhu},
\newblock \bibinfo{title}{Design of robust broadband frequency-invariant broadside beampatterns for the differential loudspeaker array},
\newblock \bibinfo{journal}{Applied Sciences (Switzerland)}  (\bibinfo{year}{2024}).
\bibitem[{Capon(1969)}]{capon1969high}
\bibinfo{author}{J.~Capon},
\newblock \bibinfo{title}{High-resolution frequency-wavenumber spectrum analysis},
\newblock \bibinfo{journal}{Proceedings of the IEEE} \bibinfo{volume}{57} (\bibinfo{year}{1969}) \bibinfo{pages}{1408--1418}.
\bibitem[{Frost(1972)}]{frost1972algorithm}
\bibinfo{author}{O.~L. Frost},
\newblock \bibinfo{title}{An algorithm for linearly constrained adaptive array processing},
\newblock \bibinfo{journal}{Proceedings of the IEEE} \bibinfo{volume}{60} (\bibinfo{year}{1972}) \bibinfo{pages}{926--935}.
\bibitem[{Griffiths and Jim(1982)}]{griffiths1982alternative}
\bibinfo{author}{L.~Griffiths}, \bibinfo{author}{C.~Jim},
\newblock \bibinfo{title}{An alternative approach to linearly constrained adaptive beamforming},
\newblock \bibinfo{journal}{IEEE Transactions on antennas and propagation} \bibinfo{volume}{30} (\bibinfo{year}{1982}) \bibinfo{pages}{27--34}.
\bibitem[{Cox et~al.(1987)Cox, Zeskind, and Owen}]{cox1987robust}
\bibinfo{author}{H.~Cox}, \bibinfo{author}{R.~Zeskind}, \bibinfo{author}{M.~Owen},
\newblock \bibinfo{title}{Robust adaptive beamforming},
\newblock \bibinfo{journal}{IEEE Transactions on Acoustics, Speech, and Signal Processing} \bibinfo{volume}{35} (\bibinfo{year}{1987}) \bibinfo{pages}{1365--1376}.
\bibitem[{Doclo and Moonen(2007)}]{doclo2007superdirective}
\bibinfo{author}{S.~Doclo}, \bibinfo{author}{M.~Moonen},
\newblock \bibinfo{title}{Superdirective beamforming robust against microphone mismatch},
\newblock \bibinfo{journal}{IEEE Transactions on Audio, Speech, and Language Processing} \bibinfo{volume}{15} (\bibinfo{year}{2007}) \bibinfo{pages}{617--631}.
\bibitem[{Wang et~al.(2023{\natexlab{a}})Wang, Yang, and Yang}]{wang2023general}
\bibinfo{author}{J.~Wang}, \bibinfo{author}{F.~Yang}, \bibinfo{author}{J.~Yang},
\newblock \bibinfo{title}{A general approach to the design of the fractional-order superdirective beamformer},
\newblock \bibinfo{journal}{IEEE Transactions on Circuits and Systems II: Express Briefs} \bibinfo{volume}{70} (\bibinfo{year}{2023}{\natexlab{a}}) \bibinfo{pages}{4291--4295}.
\bibitem[{Wang et~al.(2023{\natexlab{b}})Wang, Li, Yang, and Yang}]{wang2023robust}
\bibinfo{author}{Y.~Wang}, \bibinfo{author}{X.~Li}, \bibinfo{author}{L.~Yang}, \bibinfo{author}{Y.~Yang},
\newblock \bibinfo{title}{Robust superdirective beamforming for arbitrary sensor arrays},
\newblock \bibinfo{journal}{Applied Acoustics} \bibinfo{volume}{210} (\bibinfo{year}{2023}{\natexlab{b}}) \bibinfo{pages}{109462}.
\bibitem[{Elko(2000)}]{elko2000superdirectional}
\bibinfo{author}{G.~W. Elko},
\newblock \bibinfo{title}{Superdirectional microphone arrays},
\newblock \bibinfo{journal}{Acoustic signal processing for telecommunication}  (\bibinfo{year}{2000}) \bibinfo{pages}{181--237}.
\bibitem[{Benesty and Chen(2012)}]{benesty2012study}
\bibinfo{author}{J.~Benesty}, \bibinfo{author}{J.~Chen}, \bibinfo{title}{Study and design of differential microphone arrays}, volume~\bibinfo{volume}{6}, \bibinfo{publisher}{Springer Science \& Business Media}, \bibinfo{year}{2012}.
\bibitem[{Benesty et~al.(2015)Benesty, Chen, and Cohen}]{benesty2015design}
\bibinfo{author}{J.~Benesty}, \bibinfo{author}{J.~Chen}, \bibinfo{author}{I.~Cohen}, \bibinfo{title}{Design of circular differential microphone arrays}, volume~\bibinfo{volume}{12}, \bibinfo{publisher}{Springer}, \bibinfo{year}{2015}.
\bibitem[{Zheng and Zhi(2025)}]{zheng2025design}
\bibinfo{author}{P.~Zheng}, \bibinfo{author}{Y.~Zhi},
\newblock \bibinfo{title}{Design of differential microphone array beampatterns with sidelobe level constraints},
\newblock \bibinfo{journal}{Signal Processing} \bibinfo{volume}{227} (\bibinfo{year}{2025}) \bibinfo{pages}{109748}.
\bibitem[{Jin et~al.(2025)Jin, Luo, Huang, Chen, and Benesty}]{jin2025design}
\bibinfo{author}{J.~Jin}, \bibinfo{author}{X.~Luo}, \bibinfo{author}{G.~Huang}, \bibinfo{author}{J.~Chen}, \bibinfo{author}{J.~Benesty},
\newblock \bibinfo{title}{On the design of robust linear differential microphone arrays against low-rank noise},
\newblock \bibinfo{journal}{IEEE Transactions on Audio, Speech and Language Processing}  (\bibinfo{year}{2025}).
\bibitem[{Chen et~al.(2014)Chen, Benesty, and Pan}]{chen2014design}
\bibinfo{author}{J.~Chen}, \bibinfo{author}{J.~Benesty}, \bibinfo{author}{C.~Pan},
\newblock \bibinfo{title}{On the design and implementation of linear differential microphone arrays},
\newblock \bibinfo{journal}{The Journal of the Acoustical Society of America} \bibinfo{volume}{136} (\bibinfo{year}{2014}) \bibinfo{pages}{3097--3113}.
\bibitem[{Pan et~al.(2015)Pan, Chen, and Benesty}]{pan2015theoretical}
\bibinfo{author}{C.~Pan}, \bibinfo{author}{J.~Chen}, \bibinfo{author}{J.~Benesty},
\newblock \bibinfo{title}{Theoretical analysis of differential microphone array beamforming and an improved solution},
\newblock \bibinfo{journal}{IEEE/ACM Transactions on Audio, Speech, and Language Processing} \bibinfo{volume}{23} (\bibinfo{year}{2015}) \bibinfo{pages}{2093--2105}.
\bibitem[{Tu and Chen(2019)}]{tu2019mainlobe}
\bibinfo{author}{Q.~Tu}, \bibinfo{author}{H.~Chen},
\newblock \bibinfo{title}{On mainlobe orientation of the first-and second-order differential microphone arrays},
\newblock \bibinfo{journal}{IEEE/ACM Transactions on Audio, Speech, and Language Processing} \bibinfo{volume}{27} (\bibinfo{year}{2019}) \bibinfo{pages}{2025--2040}.
\bibitem[{Huang et~al.(2017)Huang, Benesty, and Chen}]{huang2017design}
\bibinfo{author}{G.~Huang}, \bibinfo{author}{J.~Benesty}, \bibinfo{author}{J.~Chen},
\newblock \bibinfo{title}{On the design of frequency-invariant beampatterns with uniform circular microphone arrays},
\newblock \bibinfo{journal}{IEEE/ACM Transactions on Audio, Speech, and Language Processing} \bibinfo{volume}{25} (\bibinfo{year}{2017}) \bibinfo{pages}{1140--1153}.
\bibitem[{Huang et~al.(2018)Huang, Chen, and Benesty}]{huang2018insights}
\bibinfo{author}{G.~Huang}, \bibinfo{author}{J.~Chen}, \bibinfo{author}{J.~Benesty},
\newblock \bibinfo{title}{Insights into frequency-invariant beamforming with concentric circular microphone arrays},
\newblock \bibinfo{journal}{IEEE/ACM Transactions on Audio, Speech, and Language Processing} \bibinfo{volume}{26} (\bibinfo{year}{2018}) \bibinfo{pages}{2305--2318}.
\bibitem[{Wang et~al.(2023)Wang, Yang, Li, Sun, and Yang}]{wang2023mode}
\bibinfo{author}{J.~Wang}, \bibinfo{author}{F.~Yang}, \bibinfo{author}{J.~Li}, \bibinfo{author}{H.~Sun}, \bibinfo{author}{J.~Yang},
\newblock \bibinfo{title}{Mode matching-based beamforming with frequency-wise truncation order for concentric circular differential microphone arrays},
\newblock \bibinfo{journal}{The Journal of the Acoustical Society of America} \bibinfo{volume}{154} (\bibinfo{year}{2023}) \bibinfo{pages}{3931--3940}.
\bibitem[{Zhao et~al.(2024)Zhao, Luo, Huang, Chen, and Benesty}]{zhao2024differential}
\bibinfo{author}{X.~Zhao}, \bibinfo{author}{X.~Luo}, \bibinfo{author}{G.~Huang}, \bibinfo{author}{J.~Chen}, \bibinfo{author}{J.~Benesty},
\newblock \bibinfo{title}{Differential beamforming with null constraints for spherical microphone arrays},
\newblock in: \bibinfo{booktitle}{ICASSP 2024-2024 IEEE International Conference on Acoustics, Speech and Signal Processing (ICASSP)}, \bibinfo{organization}{IEEE}, \bibinfo{year}{2024}, pp. \bibinfo{pages}{776--780}.
\bibitem[{Itzhak et~al.(2022)Itzhak, Benesty, and Cohen}]{itzhak2022multistage}
\bibinfo{author}{G.~Itzhak}, \bibinfo{author}{J.~Benesty}, \bibinfo{author}{I.~Cohen},
\newblock \bibinfo{title}{Multistage approach for steerable differential beamforming with rectangular arrays},
\newblock \bibinfo{journal}{Speech Communication} \bibinfo{volume}{142} (\bibinfo{year}{2022}) \bibinfo{pages}{61--76}.
\bibitem[{Zhao et~al.(2023)Zhao, Huang, Chen, and Benesty}]{zhao2023design}
\bibinfo{author}{X.~Zhao}, \bibinfo{author}{G.~Huang}, \bibinfo{author}{J.~Chen}, \bibinfo{author}{J.~Benesty},
\newblock \bibinfo{title}{Design of 2d and 3d differential microphone arrays with a multistage framework},
\newblock \bibinfo{journal}{IEEE/ACM Transactions on Audio, Speech, and Language Processing} \bibinfo{volume}{31} (\bibinfo{year}{2023}) \bibinfo{pages}{2016--2031}.
\bibitem[{Huang et~al.(2020)Huang, Cohen, Chen, and Benesty}]{huang2020continuously}
\bibinfo{author}{G.~Huang}, \bibinfo{author}{I.~Cohen}, \bibinfo{author}{J.~Chen}, \bibinfo{author}{J.~Benesty},
\newblock \bibinfo{title}{Continuously steerable differential beamformers with null constraints for circular microphone arrays},
\newblock \bibinfo{journal}{The Journal of the Acoustical Society of America} \bibinfo{volume}{148} (\bibinfo{year}{2020}) \bibinfo{pages}{1248--1258}.
\bibitem[{Wang et~al.(2021)Wang, Huang, Cohen, Benesty, and Chen}]{wang2021robust}
\bibinfo{author}{X.~Wang}, \bibinfo{author}{G.~Huang}, \bibinfo{author}{I.~Cohen}, \bibinfo{author}{J.~Benesty}, \bibinfo{author}{J.~Chen},
\newblock \bibinfo{title}{Robust steerable differential beamformers with null constraints for concentric circular microphone arrays},
\newblock in: \bibinfo{booktitle}{ICASSP 2021-2021 IEEE International Conference on Acoustics, Speech and Signal Processing (ICASSP)}, \bibinfo{organization}{IEEE}, \bibinfo{year}{2021}, pp. \bibinfo{pages}{4465--4469}.
\bibitem[{Zhao et~al.(2014)Zhao, Benesty, and Chen}]{zhao2014design}
\bibinfo{author}{L.~Zhao}, \bibinfo{author}{J.~Benesty}, \bibinfo{author}{J.~Chen},
\newblock \bibinfo{title}{Design of robust differential microphone arrays},
\newblock \bibinfo{journal}{IEEE/ACM Transactions on Audio, Speech, and Language Processing} \bibinfo{volume}{22} (\bibinfo{year}{2014}) \bibinfo{pages}{1455--1466}.
\bibitem[{Zhao et~al.(2016)Zhao, Benesty, and Chen}]{zhao2016design}
\bibinfo{author}{L.~Zhao}, \bibinfo{author}{J.~Benesty}, \bibinfo{author}{J.~Chen},
\newblock \bibinfo{title}{Design of robust differential microphone arrays with the jacobi--anger expansion},
\newblock \bibinfo{journal}{Applied Acoustics} \bibinfo{volume}{110} (\bibinfo{year}{2016}) \bibinfo{pages}{194--206}.
\bibitem[{Wang et~al.(2024)Wang, Yang, Hu, and Yang}]{wang2024theoretical}
\bibinfo{author}{J.~Wang}, \bibinfo{author}{F.~Yang}, \bibinfo{author}{X.~Hu}, \bibinfo{author}{J.~Yang},
\newblock \bibinfo{title}{Theoretical analysis of maclaurin expansion based linear differential microphone arrays and improved solutions},
\newblock \bibinfo{journal}{IEEE/ACM Transactions on Audio, Speech, and Language Processing}  (\bibinfo{year}{2024}).
\bibitem[{Huang et~al.(2022)Huang, Benesty, and Chen}]{huang2022fundamental}
\bibinfo{author}{G.~Huang}, \bibinfo{author}{J.~Benesty}, \bibinfo{author}{J.~Chen},
\newblock \bibinfo{title}{Fundamental approaches to robust differential beamforming with high directivity factors},
\newblock \bibinfo{journal}{IEEE/ACM Transactions on Audio, Speech, and Language Processing} \bibinfo{volume}{30} (\bibinfo{year}{2022}) \bibinfo{pages}{3074--3088}.
\bibitem[{Hao et~al.(2023)Hao, Wang, Zhang, and Yang}]{hao2023optimization}
\bibinfo{author}{X.~Hao}, \bibinfo{author}{Y.~Wang}, \bibinfo{author}{Y.~Zhang}, \bibinfo{author}{Y.~Yang},
\newblock \bibinfo{title}{An optimization method for frequency-invariant beamforming with arbitrary sensor arrays},
\newblock \bibinfo{journal}{Applied Acoustics} \bibinfo{volume}{207} (\bibinfo{year}{2023}) \bibinfo{pages}{109328}.
\bibitem[{Xie et~al.(2024)Xie, Zhao, Zhang, Benesty, and Chen}]{xie2024design}
\bibinfo{author}{J.~Xie}, \bibinfo{author}{X.~Zhao}, \bibinfo{author}{J.~Zhang}, \bibinfo{author}{J.~Benesty}, \bibinfo{author}{J.~Chen},
\newblock \bibinfo{title}{On the design of robust differential beamformers from the beampattern error perspective},
\newblock \bibinfo{journal}{IEEE Signal Processing Letters}  (\bibinfo{year}{2024}).
\bibitem[{Bi et~al.(2025)Bi, Yang, Li, Shen, and Yang}]{bi2025design}
\bibinfo{author}{H.~Bi}, \bibinfo{author}{H.~Yang}, \bibinfo{author}{L.~Li}, \bibinfo{author}{M.~Shen}, \bibinfo{author}{S.~Yang},
\newblock \bibinfo{title}{Design of a robust steerable differential beamformer with linear acoustic vector sensor arrays},
\newblock \bibinfo{journal}{Digital Signal Processing} \bibinfo{volume}{158} (\bibinfo{year}{2025}) \bibinfo{pages}{104949}.
\bibitem[{Jin et~al.(2020)Jin, Huang, Wang, Chen, Benesty, and Cohen}]{jin2020steering}
\bibinfo{author}{J.~Jin}, \bibinfo{author}{G.~Huang}, \bibinfo{author}{X.~Wang}, \bibinfo{author}{J.~Chen}, \bibinfo{author}{J.~Benesty}, \bibinfo{author}{I.~Cohen},
\newblock \bibinfo{title}{Steering study of linear differential microphone arrays},
\newblock \bibinfo{journal}{IEEE/ACM Transactions on Audio, Speech, and Language Processing} \bibinfo{volume}{29} (\bibinfo{year}{2020}) \bibinfo{pages}{158--170}.
\bibitem[{Yu(2023)}]{yu2023eigenbeam}
\bibinfo{author}{G.~Yu},
\newblock \bibinfo{title}{Eigenbeam-space transformation based steerable differential beamforming for linear arrays},
\newblock \bibinfo{journal}{Signal Processing} \bibinfo{volume}{212} (\bibinfo{year}{2023}) \bibinfo{pages}{109171}.
\bibitem[{Zhang et~al.(2023)Zhang, Mao, Cai, Ye, and Zhu}]{zhang2023broadband}
\bibinfo{author}{Y.~Zhang}, \bibinfo{author}{J.~Mao}, \bibinfo{author}{Y.~Cai}, \bibinfo{author}{C.~Ye}, \bibinfo{author}{Q.~Zhu},
\newblock \bibinfo{title}{Broadband frequency-invariant broadside beamforming with a differential loudspeaker array},
\newblock in: \bibinfo{booktitle}{2023 31st European Signal Processing Conference (EUSIPCO)}, \bibinfo{organization}{IEEE}, \bibinfo{year}{2023}, pp. \bibinfo{pages}{1728--1732}.
\bibitem[{Zhang et~al.(2024)Zhang, Xiang, and Zhu}]{zhang2024design}
\bibinfo{author}{Y.~Zhang}, \bibinfo{author}{Q.~Xiang}, \bibinfo{author}{Q.~Zhu},
\newblock \bibinfo{title}{Design of differential loudspeaker line array for steerable frequency-invariant beamforming},
\newblock \bibinfo{journal}{Sensors (Basel, Switzerland)} \bibinfo{volume}{24} (\bibinfo{year}{2024}) \bibinfo{pages}{6277}.
\bibitem[{Luo et~al.(2023)Luo, Jin, Huang, Chen, and Benesty}]{luo2023design}
\bibinfo{author}{X.~Luo}, \bibinfo{author}{J.~Jin}, \bibinfo{author}{G.~Huang}, \bibinfo{author}{J.~Chen}, \bibinfo{author}{J.~Benesty},
\newblock \bibinfo{title}{Design of steerable linear differential microphone arrays with omnidirectional and bidirectional sensors},
\newblock \bibinfo{journal}{IEEE Signal Processing Letters} \bibinfo{volume}{30} (\bibinfo{year}{2023}) \bibinfo{pages}{463--467}.
\bibitem[{Luo et~al.(2024)Luo, Jin, Huang, Chen, and Benesty}]{luo2024design}
\bibinfo{author}{X.~Luo}, \bibinfo{author}{J.~Jin}, \bibinfo{author}{G.~Huang}, \bibinfo{author}{J.~Chen}, \bibinfo{author}{J.~Benesty},
\newblock \bibinfo{title}{Design of fully steerable differential beamformers with linear superarrays},
\newblock \bibinfo{journal}{IEEE/ACM Transactions on Audio, Speech, and Language Processing}  (\bibinfo{year}{2024}).
\bibitem[{Grant et~al.(2009)Grant, Boyd, and Ye}]{grant2009cvx}
\bibinfo{author}{M.~Grant}, \bibinfo{author}{S.~Boyd}, \bibinfo{author}{Y.~Ye}, \bibinfo{title}{Cvx users’ guide}, \bibinfo{howpublished}{Online: \url{http://www.stanford.edu/boyd/software.html}}, \bibinfo{year}{2009}.

\end{thebibliography}

\end{document}